\documentclass{llncs}

\usepackage{graphicx}%
\usepackage{multirow}%
\usepackage{amsmath,amssymb,amsfonts}%
\usepackage{mathrsfs}%
\usepackage[title]{appendix}%
\usepackage{xcolor}%
\usepackage{textcomp}%
\usepackage{manyfoot}%
\usepackage{booktabs}%
\usepackage{algorithm}%
\usepackage{algorithmicx}%
\usepackage{algpseudocode}%
\usepackage{listings}%

\usepackage{tabularx}
\usepackage{hyperref}
\usepackage{xurl}
\usepackage{array}
\usepackage{comment}

\usepackage{tabularx}
\usepackage{booktabs}

\usepackage{comment}

\usepackage{booktabs,longtable,tabularx,array,multirow}
\usepackage{tikz}
 \usetikzlibrary{matrix} 
\usetikzlibrary{arrows.meta,positioning,shapes.geometric,shapes.symbols,shapes.misc,fit,backgrounds,calc,matrix,decorations.pathreplacing}
\usepackage{pdflscape}
\usepackage{caption}
\usepackage{subcaption}
\usepackage{booktabs}
\usepackage{xcolor}

\usepackage{float}

\definecolor{arismaBlue}{RGB}{38,99,171}
\definecolor{arismaLightBlue}{RGB}{229,241,255}
\definecolor{arismaGreen}{RGB}{43,137,88}
\definecolor{arismaLightGreen}{RGB}{232,247,239}
\definecolor{arismaOrange}{RGB}{217,130,30}
\definecolor{arismaLightOrange}{RGB}{255,242,224}
\definecolor{arismaRed}{RGB}{179,48,64}
\definecolor{arismaLightRed}{RGB}{255,232,236}
\definecolor{arismaPurple}{RGB}{111,75,170}
\definecolor{arismaLightPurple}{RGB}{241,236,255}
\definecolor{arismaGray}{RGB}{80,80,80}
\definecolor{arismaLightGray}{RGB}{245,245,245}

\tikzset{
  arismaArrow/.style={-{Latex[length=2.6mm,width=1.8mm]}, thick, draw=black!70},
  arismaDashed/.style={-{Latex[length=2.4mm,width=1.6mm]}, thick, dashed, draw=arismaBlue!80},
  arismaLoop/.style={-{Latex[length=2.4mm,width=1.6mm]}, thick, draw=arismaOrange!90},
  phase/.style={rectangle, rounded corners=2.2mm, draw=black!65, very thick, align=center, fill=white, minimum height=9mm, text width=34mm, font=\sffamily\small},
  phaseWide/.style={rectangle, rounded corners=2.2mm, draw=black!65, very thick, align=center, fill=white, minimum height=9mm, text width=46mm, font=\sffamily\small},
  ai/.style={rectangle, rounded corners=2.2mm, draw=arismaBlue!85!black, very thick, align=center, fill=arismaLightBlue, minimum height=8mm, text width=40mm, font=\sffamily\scriptsize},
  human/.style={rectangle, rounded corners=2.2mm, draw=arismaGreen!80!black, very thick, align=center, fill=arismaLightGreen, minimum height=8mm, text width=40mm, font=\sffamily\scriptsize},
  validate/.style={diamond, aspect=2.0, draw=arismaOrange!90!black, very thick, align=center, fill=arismaLightOrange, inner sep=1.2pt, text width=25mm, font=\sffamily\scriptsize},
  risk/.style={rectangle, rounded corners=2.2mm, draw=arismaRed!85!black, very thick, align=center, fill=arismaLightRed, minimum height=8mm, text width=35mm, font=\sffamily\scriptsize},
  artefact/.style={rectangle, draw=black!55, thick, align=center, fill=arismaLightGray, minimum height=7mm, text width=36mm, font=\sffamily\scriptsize},
  header/.style={font=\sffamily\bfseries\large, align=center},
  smallnote/.style={font=\sffamily\scriptsize, align=center, text=black!75},
  laneLabel/.style={font=\sffamily\bfseries\small, text=black!75, align=center},
  flowbox/.style={rectangle, rounded corners=2mm, draw=black!65, thick, align=center, fill=white, minimum height=8mm, text width=34mm, font=\sffamily\scriptsize},
  countbox/.style={rectangle, rounded corners=2mm, draw=arismaBlue!80!black, thick, align=center, fill=arismaLightBlue, minimum height=8mm, text width=40mm, font=\sffamily\scriptsize},
  controlbox/.style={rectangle, rounded corners=2mm, draw=arismaGreen!80!black, thick, align=center, fill=arismaLightGreen, minimum height=8mm, text width=40mm, font=\sffamily\scriptsize},
  matrixcell/.style={rectangle, rounded corners=1mm, draw=black!45, align=center, fill=white, text width=24mm, minimum height=9mm, font=\sffamily\scriptsize},
  matrixhead/.style={rectangle, rounded corners=1mm, draw=black!55, align=center, fill=black!8, text width=24mm, minimum height=9mm, font=\sffamily\bfseries\scriptsize}
}

\usepackage[T1]{fontenc}
\usepackage{graphicx}
\usepackage{booktabs}


\usepackage{tcolorbox}

\usepackage{placeins}

\usepackage{eso-pic}
\usepackage{url}
\newcommand\copyrighttext{%
\footnotesize
This manuscript has been accepted for presentation at {\bf \em AGENTICS 2026}, the International Conference on Agentic and Generative Techniques in Intelligent Computational Systems,
held in Angers, France, 28-30 October 2026, and for {\bf \em publication
in Springer proceedings}.
This is the authors' manuscript version. The final authenticated publication will be available via Springer.
}
\newcommand{\copyrightnotice}{%
\AddToShipoutPictureFG*{%
\AtPageLowerLeft{%
\raisebox{1cm}{%
\makebox[\paperwidth][c]{%
\fbox{\parbox{\textwidth}{\copyrighttext}}%
}}}}}

\hypersetup{
  hidelinks,
  pdftitle={ARISMA: Guidelines for AI- and LLM-Assisted Systematic Reviews, Scoping Reviews, and Mapping Studies},
  pdfauthor={Mahyar Tourchi Moghaddam and Mina Alipour},
  pdfkeywords={AI-assisted review, systematic review, scoping review, mapping study, large language models}
}

\begin{document}
\title{ARISMA: Guidelines for AI- and LLM-Assisted Systematic Reviews, Scoping Reviews, and Mapping Studies}

%\titlerunning{ARISMA: AI-Assisted Evidence Synthesis}

%\author{Omitted for review.}

%\author{Anonymized for review.}
%\institute{}

\author{
Mahyar Tourchi Moghaddam\orcidID{0000-0001-5028-7546}\and
Mina Alipour\orcidID{0000-0002-6717-8976}
}
\authorrunning{Moghaddam et al.}
\institute{University of Southern Denmark, Odense 5230, DK \\
\email{\{mtmo,mial\}@mmmi.sdu.dk}
}

\begin{comment}
\sur{Moghaddam}\orcid{0000-0001-5028-7546}
\email{mtmo@mmmi.sdu.dk}
\author{\fnm{Mina} \sur{Alipour}\orcid{0000-0002-6717-8976}}
\email{mial@mmmi.sdu.dk}

\affil{\orgname{University of Southern Denmark},
\street{Campusvej 55},
\city{Odense},
\postcode{5230},
\country{Denmark}}
\end{comment}

            % typeset the header of the contribution
%

\maketitle
\copyrightnotice
\begin{abstract}
Systematic reviews, scoping reviews, mapping studies, and related evidence syntheses are increasingly difficult to conduct with fully manual workflows as search volumes, update cycles, and synthesis requirements continue to expand. At the same time, artificial intelligence, machine learning, and large language models are rapidly entering review practice across query formulation, screening, extraction, categorization, appraisal support, and reporting. Yet the empirical evidence remains uneven, task-dependent, and insufficient to justify unconstrained automation. Existing standards such as PRISMA 2020, PRISMA-S, PRISMA-ScR, PRISMA-P, PRESS, and SWiM remain essential, but none provides an end-to-end operational standard for when AI use is methodologically appropriate, how it should be validated, which review decisions must remain human-led, and how AI involvement should be reported so that readers can audit it. This paper proposes ARISMA, an {\bf A}I {\bf R}eporting and {\bf I}ntegration standard for {\bf S}ystematic {\bf M}ethods and {\bf A}nalysis. ARISMA treats AI as an inspected, benchmarked, logged, and reversible assistant rather than an autonomous reviewer. It is built around one governing principle: every consequential scientific decision must remain human-interpretable, human-auditable, and human-accountable. The paper contributes a lifecycle taxonomy, process guidance, stepwise recommendations across the review pipeline, a governance and provenance model, a tool-support framework, an AI-integrated reporting checklist, and a validation matrix. It also addresses legal, privacy, infrastructure, and sustainability considerations. %The framework was iteratively refined through structured validation. 
The framework was iteratively refined through structured expert consultation. 
The result is a practical and auditable guideline for responsible AI-assisted evidence synthesis.
%\keywords{}


\end{abstract}

%\maketitle  
%\begin{thebibliography}
%\bibliography{bib}
%\end{thebibliography}

%%
\section{Introduction}

Systematic reviews do not merely summarize literature; they define what counts as the current state of the art for researchers, practitioners, policymakers, funders, and guideline developers. That privileged role explains why reporting standards such as PRISMA 2020 \cite{page2021prisma} have become foundational across disciplines: the credibility of a review depends not only on the studies it includes, but also on whether the review process itself is transparent, reproducible, and open to critical scrutiny. PRISMA provides the modern reporting anchor through a statement paper, checklist, expanded checklist, abstract checklist, and flow diagrams, while related extensions such as PRISMA-S \cite{rethlefsen2021prisma}, PRISMA-ScR \cite{tricco2018prisma}, PRISMA-P \cite{moher2015preferred}, and SWiM \cite{campbell2020synthesis} support search transparency, scoping reviews, protocols, and non-meta-analytic synthesis. Together, these guidelines have substantially improved review reporting, but they were not intended as a governance standard for AI-assisted review.

The review family itself has also expanded as many contemporary studies are not conventional intervention-effect reviews. Scoping reviews are used to clarify concepts, characterize bodies of evidence, and identify gaps \cite{campbell2023mapping}; mapping reviews and evidence-and-gap maps emphasize landscape description, categorization, and evidence distribution \cite{khalil2025methodology}; and software-engineering mapping studies often rely on keywording, classification schemes, and structured maps rather than pooled causal estimation \cite{kitchenham2013systematic}. The implication is methodologically important: a guideline for AI-assisted evidence synthesis cannot assume a single review product. The acceptable role of AI depends on whether the review aims to estimate effects, map a field, structure a taxonomy, monitor a living corpus, or support broader evidence-informed reflection.

The urgent need for an AI-aware guideline arises from the collision of two developments: {\em i)} evidence synthesis has become slower, costlier, and more difficult to keep current; {\em ii)}  AI systems are already entering the workflow, often before robust norms for validation, logging, and disclosure have stabilized. A recent cross-disciplinary review found that only about 5\% explicitly reported machine-learning use, mostly for screening, with much rarer use in search, extraction, or synthesis \cite{scotti2025artificial}. A recent scoping review of LLM use in systematic reviews identified 37 relevant studies covering 10 of 13 review steps; search, screening, and extraction were the most common targets, and results were mixed rather than uniformly positive \cite{lieberum2025large}.

There is methodological opportunity as LLM-assisted title and abstract screening has shown promising performance in some settings. In one public-policy review of opioid-related policies, GPT-4 recommended excluding 41,742 of 43,480 articles with abstracts, leaving 1,738 for manual assessment, and the validation study reported an estimated false-exclusion rate of 0.00, with an upper confidence bound of 0.05 in the final test \cite{rubinstein2025using}. In a living systematic review setting, refined GPT-4o prompts achieved 100\% sensitivity for studies ultimately included after full-text screening, with simulated workload reductions of 65\% to 85\% \cite{homiar2025development}. In a three-layer psychiatric-review study, GPT-4 reached human-comparable sensitivity with very high specificity after accounting for justified exclusions and processed roughly 110 records per minute \cite{matsui2024human}. At the same time, other evaluations found only poor-to-moderate performance for several screening and extraction tasks outside optimized settings, which is why permissive claims about {\em autonomous reviewing} are not justified by the present evidence base \cite{lieberum2025large}.

There is also methodological risk, e.g., search-string generation with LLMs can be useful for brainstorming and initial query construction, but recent evaluations show they still lack the sensitivity and precision required for unsupervised final search design \cite{adam2024literature}. Data extraction performance varies substantially across domains and variable types, and when LLMs summarize scientific research, they can overgeneralize beyond what the original studies support: research showed LLM-generated science summaries were nearly five times more likely than human summaries to contain overly broad generalizations, with some models overgeneralizing in 26\% to 73\% of cases \cite{peters2025generalization}.

That creates the central design challenge for a modern guideline paper: {\em how to use AI without surrendering scientific control}. Researchers \cite{flemyng2025position} emphasize that evidence synthesists remain ultimately responsible for the review, that AI must be used in ways consistent with legal and ethical standards, and that transparent reporting is mandatory. The same direction is reinforced by several publishers that require or strongly recommend disclosure of AI use and reject the idea that AI systems can bear authorship responsibility.

%That problem is not adequately solved by existing guidance.
Existing reporting standards remain essential but incomplete for AI-assisted review. 
PRISMA 2020 \cite{page2021prisma} instructs authors on what to report in a systematic review; PRISMA-S \cite{rethlefsen2021prisma} clarifies how to report searches; PRISMA-P \cite{moher2015preferred} supports protocol registration; PRISMA-ScR \cite{tricco2018prisma} supports scoping-review reporting; PRESS \cite{mcgowan2016press} structures peer review of search strategies; and SWiM \cite{campbell2020synthesis} helps reviewers report synthesis when meta-analysis is not appropriate. While remaining indispensable, none of them answers five questions that have become central in AI-assisted review practice: {\em i)} when AI use is methodologically justified; {\em ii)} which tasks are low-risk enough for assistive automation; {\em iii)} which outputs require benchmark validation before they can influence the evidence base; {\em iv)} which decisions require mandatory human sign-off; and {\em v)} what exactly readers must be told if AI has shaped the search, screening, extraction, coding, or prose. ARISMA is designed to fill that gap without displacing the classical standards on which it depends.

This gap is not only methodological but ethical: AI-assisted evidence synthesis redistributes epistemic authority across reviewers, models, vendors, platforms, and infrastructures. If ungoverned, that redistribution can affect inclusion decisions, evidence visibility, interpretive framing, privacy, accountability, and the credibility of downstream policy or clinical guidance. ARISMA therefore treats AI use in evidence synthesis as an ethics-of-methods problem: the central question is not whether AI can accelerate review work, but under what governance conditions such acceleration remains transparent, accountable, privacy-respecting, and scientifically justified.

%The position taken in this paper is intentionally conservative. 
ARISMA does not treat AI as a replacement reviewer, but as an inspected, bounded, and reversible assistant. In ARISMA, the role of AI is acceptable only when four conditions are met: {\em i)} the task is explicitly delimited; {\em ii)} the model’s inputs, outputs, and constraints are documented; {\em iii)} performance has been validated against a human reference appropriate to the task; and {\em iv)} humans retain both override power and accountability for all consequential decisions. This framing aligns with the current direction of AI risk-management thinking, which emphasizes trustworthiness, governance, transparency, validation, and documented human responsibility rather than unqualified automation \cite{nist}. 

ARISMA is designed as a governance and reporting framework rather than as a performance claim about any individual model, vendor, platform, or automation architecture. The framework therefore focuses on methodological accountability, provenance, validation, auditability, and human oversight principles that remain applicable despite rapid evolution in AI ecosystems.

This paper makes four contributions:

\begin{itemize}
    \item It offers an end-to-end taxonomy of AI roles across systematic reviews, scoping reviews, and mapping studies.

    \item It provides stepwise methodological guidance for each major review activity, including searching, deduplication, screening, snowballing, extraction, categorization, synthesis, and reporting.

    \item It adds a governance layer through explicit validation workflows, provenance tracking, audit requirements, and human checkpoints.

    %\item It adds a governance layer through validated workflows, provenance tracking, audit requirements, and explicit human checkpoints. 

    \item It introduces practical reporting instruments, an AI-integrated checklist, validation matrix, and AI-aware evidence-flow logic that allow editors, peer reviewers, and readers to inspect where AI entered the process and whether its role remained scientifically legitimate.

\end{itemize}

In this way, ARISMA seeks to make AI-assisted reviews more auditable, more governable, and more defensible rather than making reviews {\em more automatic}.

%The remainder of the paper proceeds from design logic to operational use. Section 2 introduces the ARISMA lifecycle, explains the taxonomy of AI roles, and proposes the review workflow. Section 3 develops the governance layer, including validation, provenance, legal and privacy safeguards, model-agreement strategies, expert validation, and sustainability considerations. Section 4 explains bounded current tool support across review steps. Section 5 operationalizes ARISMA through the checklist, validation matrix, and AI-aware evidence-flow logic. Sections 6 and 7 then close with threats to validity and the implications of ARISMA as a conservative standard for AI-assisted evidence synthesis.

The remainder of the paper proceeds from development logic to operational use. %Section 2 describes the development basis and expert validation of ARISMA. 
Section 2 describes the development basis and expert-informed refinement of ARISMA. 
Section 3 introduces the framework and operational process. Section 4 develops the governance, validation, and provenance layer. Section 5 reviews current tool support and sustainability considerations. Section 6 presents the reporting checklist and validation matrix. Section 7 discusses threats to validity, Section 8 presents declarations, and Section 9 concludes with implications for responsible AI-assisted evidence synthesis.

\section{Development Basis and Expert-Informed Refinement of ARISMA}

{\bf Development rationale.}
ARISMA was developed through a literature-informed design process that
combined established evidence-synthesis guidance, recent empirical and
methodological work on AI-assisted review methods, and structured expert
consultation. The authors first identified methodological commitments shared
across major review standards, including transparent search reporting,
explicit eligibility criteria, reproducible selection procedures, traceable
extraction, and defensible synthesis
\cite{page2021prisma}. They then examined where AI
systems were being introduced into these activities
\cite{lieberum2025large}.
This process produced a draft lifecycle model, an operating framework, and
initial reporting and validation instruments. The preliminary framework was
intentionally conservative, with emphasis on auditability, reversibility, and
human accountability.

{\bf Structured expert consultation.}
The draft framework was reviewed through 45- to 60-minute video consultations
with 21 researchers and information specialists experienced in evidence
synthesis. Participants were approached purposively through recent
methodological and AI-assisted evidence-synthesis publications, with
additional recommendations from initial consultees. The purposive approach
was selected to obtain relevant methodological expertise rather than a
statistically representative sample \cite{robson2024real}. The semi-structured consultation format was informed by problem-centered expert interviewing \cite{doringer2021problem}. Discussion focused on five
areas: {\em i)} lifecycle coverage; {\em ii)} the boundaries of assistive, adjudicative, and
generative AI use; {\em iii)} the clarity of governance checkpoints; {\em iv)} the usability of
the checklist and validation matrix; and {\em v)} the accuracy of the figures.

{\bf Review and use of feedback.}
The authors organized the consultation feedback around these five areas and
used the issues raised to guide revision. The exercise was formative: it was
intended to identify omissions and improve the clarity and practical
usability of ARISMA. Key issues raised concerned benchmark validation, model-version and prompt logging, model instability, stopping
rules, and human oversight for high-consequence tasks.

{\bf Resulting refinements.}
In response to the consultation, the calibration and pilot-testing
requirements were expanded; model metadata and prompt documentation were
added to the reporting checklist; conditions for revising or disabling AI
support were clarified; and responsibilities for human review of screening,
extraction, appraisal, and synthesis outputs were strengthened. These
refinements were considered together with the published methodological and
empirical literature, which remains the principal evidential basis for
ARISMA.

{\bf Consent and confidentiality.}
All experts provided informed consent for participation and for the use of
anonymized methodological feedback. Identifying information was removed from
the consultation records used during manuscript preparation, and no
identifiable quotations are reported.

\section{ARISMA Framework and Operational Process}

\subsection{Taxonomy and AI-use categories}

ARISMA organizes AI-assisted evidence synthesis into six linked phases of {\bf \em framing, protocolization, retrieval, selection and enrichment, evidence structuring,} and {\bf \em synthesis and reporting}. A taxonomy is necessary as different review families require different evidentiary products, and AI can only be judged appropriately in relation to those products. A scoping review may legitimately prioritize breadth, coding, and gap identification; a mapping study may prioritize category schemes and frequency distributions; a systematic review of effects may require outcome-level extraction, risk-of-bias assessment, and quantitative or SWiM-compliant synthesis. The acceptable role of AI changes across those designs \cite{peters2015guidance}. 

ARISMA also distinguishes assistive, adjudicative, and generative AI use. Assistive use means query suggestion, deduplication, ranking, tagging, or form prepopulation. Adjudicative use means classification, eligibility judgment, or appraisal support. Generative use means drafting text, summaries, category labels, or interpretations. The higher the epistemic consequence of the task, the stricter the validation and human oversight must be. In ARISMA, assistive uses are generally easier to justify; adjudicative uses require benchmarked performance against a human reference standard; and generative uses are never accepted as evidence in themselves, only as drafts grounded in already verified review data. That hierarchy follows directly from both the positive validation studies and the growing evidence of overgeneralization and instability in unconstrained generation \cite{peters2025generalization}.

\begin{figure}%[H]
    \centering
    \includegraphics[width=\linewidth]{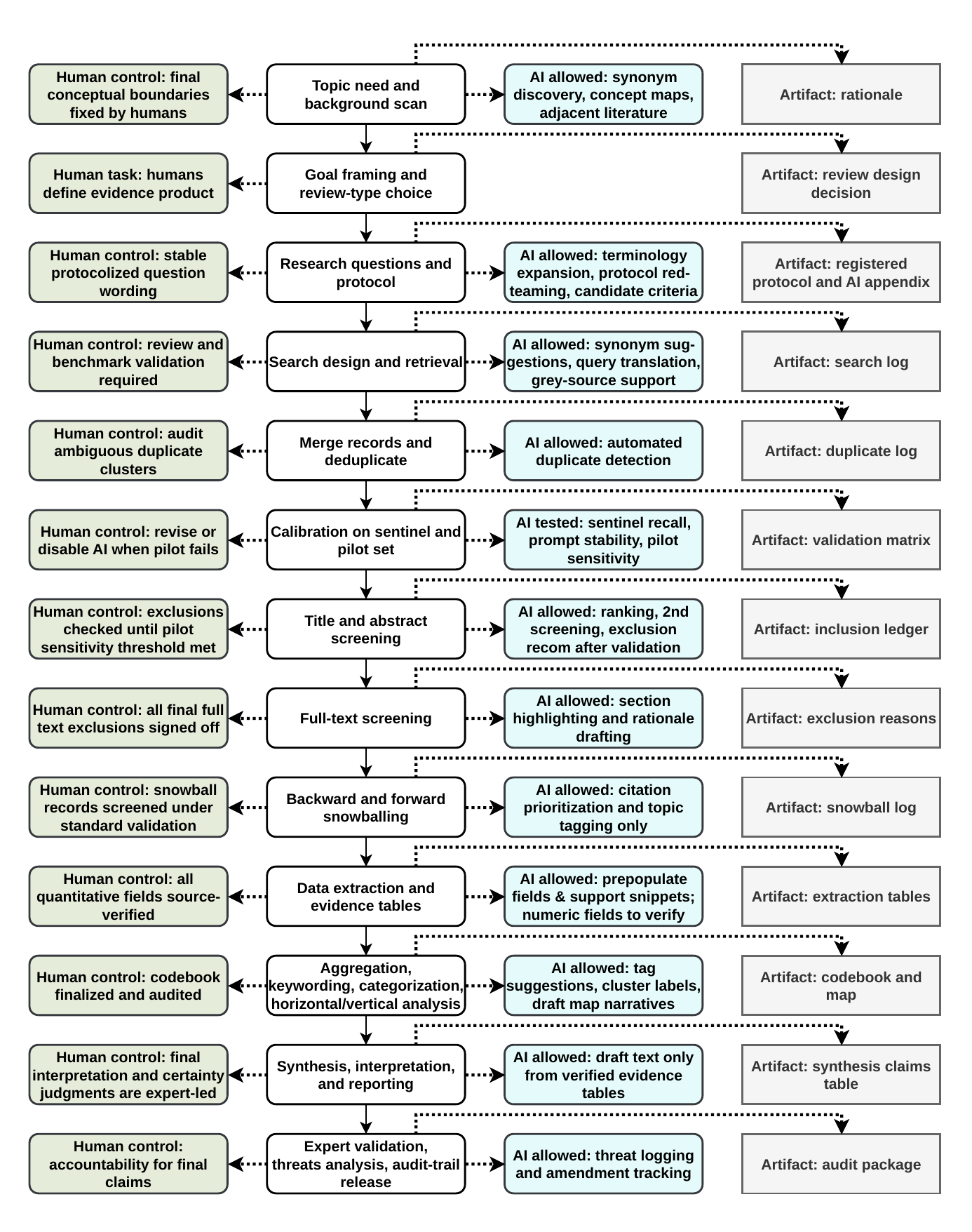}
    \caption{General ARISMA process. AI support is attached to auditable artefacts and constrained by human validation checkpoints.}
    \label{fig:arisma-general-process}
\end{figure}

Figure \ref{fig:arisma-general-process}
demonstrates the governing logic of ARISMA. Its center column (white colored) represents the review process, from background framing to expert validation and release of the final audit package. The left (green) column defines the points at which human control is mandatory, and the right (gray) column defines the auditable artefacts that each step must produce. The shaded AI boxes (in light blue) represent limited intervention zones whose legitimacy depends on human oversight and the audit trail left behind.
%are therefore not {\em automation opportunities}, but they are bounded intervention zones whose legitimacy depends on what the human reviewers allow the model to do and what trace is left behind. %This framing is crucial because the same technical model behavior can be methodologically acceptable in one location and unacceptable in another. For example, AI-generated synonym expansion may be acceptable during exploratory search design if it is benchmark-tested and manually reviewed, whereas AI-generated full-text exclusion is not acceptable unless the final judgment remains distinctly human-led.

Consultation feedback highlighted the importance of reversible AI
intervention points. Accordingly, the left column was labeled ``Human
control,'' and arrows were added to show that each AI box is a bounded
intervention requiring mandatory sign-off.

Below, we explain how this control logic applies across the review lifecycle, identifying where AI may assist, where validation is required, and where human accountability must remain explicit.

{\bf Background.} 
The background step should establish why a review is needed, what decision problem it serves, and why the chosen review type is the right design. %\cite{arksey2005scoping}.
For scoping and mapping reviews, the background must justify breadth, concept clarification, and gap identification; for effect reviews, it must justify a more focused causal or evaluative question. AI can assist by surfacing terminology variants, identifying adjacent literature, clustering early search results, and drafting concept maps. It should not determine the problem statement. A strong AI-assisted background section, therefore, uses AI only for exploratory breadth, then freezes the final conceptual boundaries in human-authored form. In practice, the safest pattern is to ask an LLM to generate alternative framings and likely synonym families, then compare those suggestions against sentinel papers, prior reviews, and domain-expert feedback before the protocol is finalized.

{\bf Goal framing.} 
The next step is to decide whether the study is a systematic review, scoping review, mapping review, evidence and gap map, or hybrid design. This decision determines not only the research question but also the acceptable role of AI. For example, AI-based cluster labeling may be very useful in a mapping review, but the same functionality is insufficient for an intervention review that will support guideline recommendations. Goal framing should therefore specify the intended output: pooled effect estimate, narrative synthesis, evidence map, taxonomy, trend analysis, gap map, or methodological landscape. If a team cannot state the final evidence product clearly, it is too early to introduce AI into any consequential step.

{\bf Research questions.} 
AI- and LLM-assisted reviews still begin with classical review-question logic. Population, Intervention, Comparison, Outcome (PICO) remains central for many effect reviews \cite{eriksen2018impact}, while Population, Concept, Context (PCC) \cite{peters2020updated} and related frameworks are common in scoping reviews, and mapping studies often rely on facet-based questions such as population, method, domain, intervention family, outcome class, country, or publication venue. AI can be useful here in decomposing a broad question into searchable concept blocks, generating candidate inclusion boundaries, and stress-testing whether a question is too broad, too narrow, or internally inconsistent. However, the final question must be protocolized in stable human language because all later validation depends on it. If the review team lets question wording drift during model prompting, AI performance cannot be interpreted, and the review ceases to be auditable.

{\bf Protocolization and registration.} 
A protocol is where ARISMA becomes conservative by design. PRISMA-P \cite{moher2015preferred} exists because prospective protocols reduce selective methods drift and improve transparency, and registries such as PROSPERO exist to reduce unintended duplication and reporting bias \cite{schiavo2019prospero}. In ARISMA, the protocol must contain not only the ordinary review methods but also an AI methods appendix stating the task, model or tool, version or build, access mode, prompt frame, input data type, validation design, human oversight rule, stopping rule, failure mode, and reporting plan. This appendix should be treated as a formal methods section, not as a software footnote. If AI use changes during the review, the amendment must be dated and justified, exactly as other protocol changes should be.

{\bf Automatic search.} 
Search is one of the most tempting places to use LLMs because query construction is painstaking and terminology is heterogeneous. ARISMA supports AI in search only under a librarian-plus-benchmark model. The model may propose synonyms, controlled vocabulary candidates, excluded homonyms, and translations across databases. It may also help translate a validated search between platforms. But it must not be trusted as the final search architect without peer review and empirical checks. The PRESS 2015 guideline exists precisely because electronic search strategies need structured quality review \cite{mcgowan2016press}. Recent work on {\em literature search sandbox} showed that an LLM trained on 10,346 PROSPERO search queries produced searches with median sensitivity of 85\% and high numbers needed to read, with librarians concluding that such queries could be useful as starting points or topic-scoping aids but not without scrutiny \cite{adam2024literature}. Other work has explicitly cautioned that ChatGPT-built review searches require stringent evaluation \cite{bencze2025comparing}. %As Table XXXXX shows, Polyglot Search Translator and 2Dsearch are valuable examples of support tools because they constrain or structure parts of the search task rather than pretending to replace an expert search method. 
 ARISMA search workflow, therefore, looks as follows. {\em First}, humans define the main concepts and a small “must-find” benchmark set. {\em Second}, AI expands synonyms and controlled vocabulary candidates. {\em Third}, the search is manually curated and PRESS-reviewed. {\em Fourth}, the candidate strategy is tested against the benchmark set and iteratively repaired until recall is acceptable. {\em Fifth}, the full strategy and all AI contributions are archived. A model-generated string that misses known studies on a pilot set is not a near miss; it is a failing search strategy. That is the level of discipline required if an AI-generated search is going to influence downstream sampling.

%Retrieval in ARISMA is treated as an iterative and benchmark-validated process rather than a one-time query-construction exercise. Figure \ref{ret} summarizes the retrieval, enrichment, deduplication, and revalidation workflow.

\begin{figure}%[H]
    \centering
    \includegraphics[width=\linewidth]{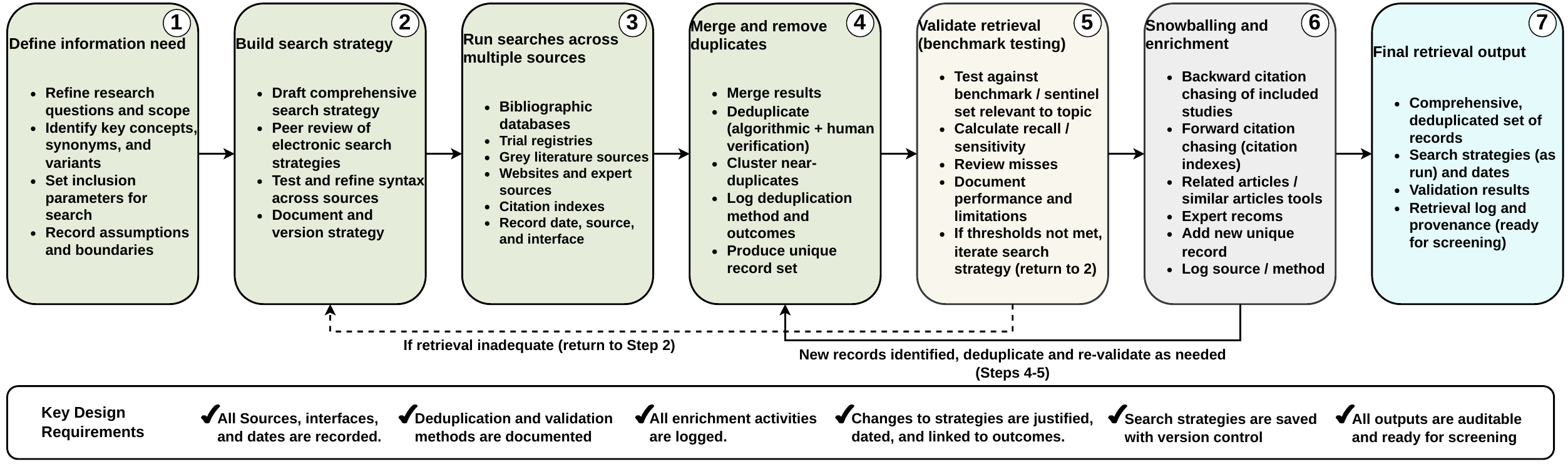}
    \caption{ARISMA Retrieval and enrichment pipeline: search, deduplication, and snowballing.}
    \label{ret}
\end{figure}

Since retrieval is where invisible downstream bias often begins, ARISMA treats search, enrichment, deduplication, and revalidation as a single controlled subsystem rather than as disconnected technical chores. We define a retrieval and enrichment pipeline as a subsystem of automatic search presented in Figure \ref{ret}. 
%makes that subsystem explicit.

%

%In conventional review descriptions, search design, database execution, duplicate removal, and snowballing are often narrated sequentially, which can obscure the fact that each stage can expose defects in earlier decisions. ARISMA instead makes the feedback structure explicit. Benchmark testing may reveal that a search string misses known sentinel studies; deduplication may reveal normalization problems inherited from source exports; snowballing may uncover terminological clusters that the original query failed to represent. The figure therefore visualizes retrieval as an iterative process whose quality is judged by recall, transparency, and provenance rather than by apparent completeness alone. 

In the defined subsystem (Figure \ref{ret}), Step 1 defines the information need and the conceptual boundaries of the review. Step 2 builds and versions the search strategy. Step 3 executes the search across multiple sources and logs interfaces, dates, and source types. Step 4 merges records and removes duplicates through a combination of algorithmic matching and human verification. Step 5 validates retrieval performance against a benchmark or sentinel set. Step 6 enriches the corpus through backward and forward snowballing and related-article expansion. Step 7 produces the final retrieval output: a comprehensive, deduplicated, provenance-aware record set ready for screening. The key design requirements shown beneath the figure are the operating conditions required for auditable retrieval and adequate governance. %If these are not met, AI-assisted retrieval has not been adequately governed. 

%ARISMA operationalizes retrieval as a loop, not a one-shot query event, since an expert noted that {\em “retrieval must be iterative rather than linear”}. The diagram includes a loop from validation back to search, emphasized as “if thresholds not met, iterate search strategy”, reminding teams that retrieval quality is judged by recall, transparency and provenance.

Consultation feedback emphasized that retrieval should be iterative rather
than treated as a one-shot query event. The diagram therefore includes a loop
from validation back to search, emphasized as ``if thresholds are not met,
iterate the search strategy,'' reminding teams that retrieval quality is
judged by recall, transparency, and provenance.

{\bf Merging and removing duplications.} 
Deduplication is not a mere cleaning task. Duplicate records distort workload estimates, can corrupt screening statistics, and in the worst case can lead to double counting if study reports are mishandled. ARISMA recommends that deduplication occur in two passes: automated duplicate detection followed by human audit of ambiguous clusters and multi-report study families. 
%ASySD is a strong current example of a task-bounded automation tool; across five unseen biomedical datasets ranging from 1,845 to 79,880 citations, ASySD identified more duplicates than EndNote or SRA-DM, with sensitivity reported at 0.95 to 0.99 and false-positive rates comparable to human performance. The methodological lesson is important: 
%Current tools (see table XXX) show that targeted automation with a clear objective and explicit evaluation is preferable to unconstrained generation. 
Current tools discussed in Section 5 reinforce the same principle: bounded automation is more defensible than unconstrained generation. 
AI or ML may therefore be trusted more in duplicate detection than in free-form summary writing, but every ambiguous duplicate family still needs a human decision \cite{hair2023automated}.

{\bf Inclusion and exclusion criteria framing.}
LLM performance in screening depends heavily on the quality and specificity of the criteria they are given. Inclusion and exclusion criteria should therefore be written as decision rules, not as aspirational prose. Criteria need explicit statements about population, exposure or intervention, outcomes, study design, language, document type, time window, publication status, and edge cases such as protocols, conference abstracts, or secondary analyses. AI is helpful here as a critic, not as the author, i.e., it can generate borderline hypotheticals, surface hidden ambiguities, and test whether different phrasings imply different judgments. Recent work in prompt development for screening and in literature prefiltering \cite{delgado2025transforming,cao2025development} concludes that clearer, better-structured criteria improve LLM screening behavior. ARISMA therefore treats {\em criteria red-teaming} as a legitimate and often useful AI contribution before actual screening begins \cite{homiar2025development}.

{\bf Applying criteria during title and abstract screening.} 
This is the most mature LLM use case, but it is not mature enough for universal autonomy. The evidence says the task can work well under the right conditions, with strong prompts, calibration, and clear asymmetry in the use case, but also that performance varies across domain, prevalence, and prompt design \cite{delgado2025transforming}. The public-policy GPT-4 feasibility study is persuasive for exclusion support in a setting where most records are irrelevant \cite{rubinstein2025using}. The living-review GPT-4o study is persuasive for prompt-refined workload reduction with continued benchmarking \cite{homiar2025development}. The three-layer GPT-4 JMIR study is persuasive for high-throughput screening under carefully engineered prompts \cite{matsui2024human}. Yet researchers show that performance remains inconsistent across tasks and languages when prompts are less optimized, or the workflow is more ambitious \cite{khraisha2024can}.
For that reason, ARISMA recommends a three-tier operating rule. In the {\em low-trust tier}, AI only ranks or labels records for human-first screening. In the {\em moderate-trust tier}, AI may recommend exclusions, but every exclusion is checked by a human until pilot sensitivity reaches a predeclared threshold against a consensus development set. In the {\em high-consequence tier}, such as reviews intended for guidelines or quantitative synthesis, AI may accelerate screening only as a second reviewer, prefilter, or prioritisation system; it must not become the sole excluding reviewer unless the team has domain-specific validation evidence and an explicit justification for doing so. This is a normative recommendation from ARISMA, grounded in the current pattern of empirical evidence rather than in any claim that a universal threshold already exists.

{\bf Full-text screening.} 
Full-text decisions carry higher consequences than title and abstract screening because they commit the review to its final evidence base. LLMs can help here by highlighting relevant sections, extracting justification snippets, and drafting include/exclude rationales, but the final judgment should remain a human consensus decision. The strongest current evidence suggests that full-text screening can approach human-like agreement only under constrained settings and highly reliable prompts, not as a general default \cite{khraisha2024can}. In ARISMA, full-text screening is therefore a human-signoff step even if an AI assistant provides section-level reasoning or document triage. A useful practice is to require the assistant to cite page- or section-level evidence for each recommendation and to reject any full-text judgment that lacks traceable support.

\begin{figure}%[H]
    \centering
    \includegraphics[width=.9\linewidth]{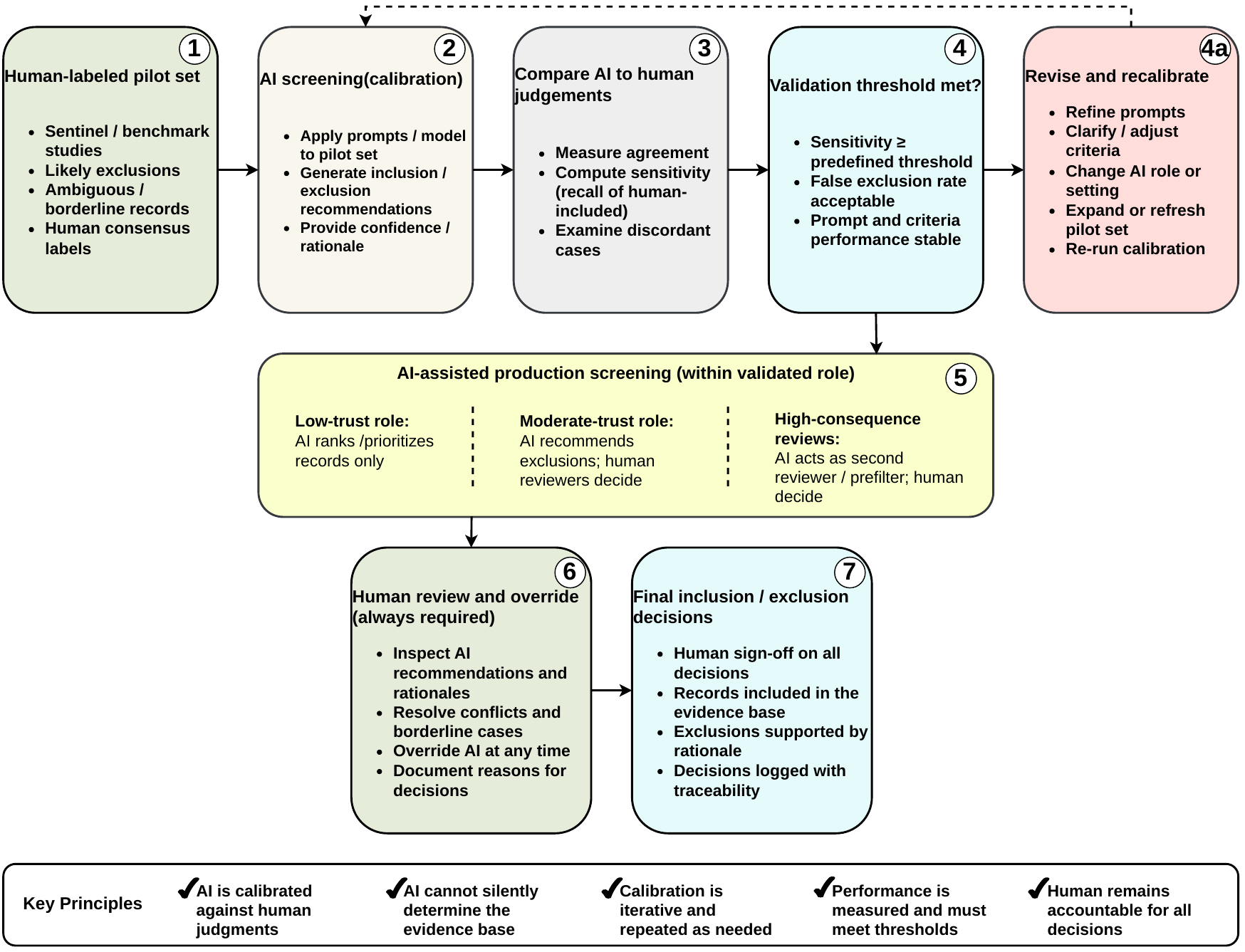}
    \caption{Screening workflow. AI is calibrated against human judgments and cannot
silently determine the evidence base.}
    \label{screen}
\end{figure}

Screening is the point at which AI can most directly and silently reshape the eventual evidence base. %Thus, Figure \ref{screen} should be explained in terms of calibration rather than automation.
%does not assume that a model may screen because it appears accurate in a generic benchmark. It
As shown in Figure \ref{screen}, ARISMA requires task-specific calibration against a human-labeled pilot set that includes sentinel inclusions, likely exclusions, and genuinely ambiguous borderline records. 
%Experts suggested splitting the workflow into a calibration phase and a production phase. 
Consultation feedback supported separating the workflow into a calibration phase and a production phase.  
The figure’s top half therefore represents a calibration experiment, not a production workflow. 
The figure shows that AI deployment should be contingent on task-specific validation through pilot testing and that, depending on the achieved sensitivity, models can be used for ranking, recommending or second review only.
Thus, AI output is first compared with human judgments, discordant cases are examined, and only then is a trust tier assigned to the model’s role. 

The lower half of Figure \ref{screen} is a tiered-permission system. In the low-trust role, AI only ranks or prioritizes records for human-first review. In the moderate-trust role, AI may recommend exclusions, but human reviewers still decide whether those exclusions stand. In high-consequence reviews (for example, reviews intended to support formal recommendations or synthesis-critical inference) AI may at most operate as a second reviewer, prefilter, or prioritisation aid, with humans retaining the exclusion decision. The important methodological point is that these tiers are not fixed properties of any model. They are conditional permissions earned through calibration, contingent on task, corpus, and documented performance thresholds. 
Figure \ref{screen} is designed asymmetrically: moving from calibration to production requires passing explicit thresholds, but failing thresholds returns the process to revision and recalibration. That asymmetry encodes a precautionary principle. In ARISMA, the burden of proof lies with AI use, not with human reviewers who object to it. %This is consistent with current risk-management thinking and with the journal’s emphasis on human accountability for substantively AI-assisted scholarly output.

{\bf Snowballing.} 
Snowballing remains essential because database searching alone can miss semantically related but terminologically disconnected studies. Wohlin’s guideline paper \cite{wohlin2014guidelines} treats backward and forward snowballing as a systematic search procedure rather than an informal afterthought and emphasizes the importance of a diverse starting set. 
%Citationchaser operationalizes this logic in a reproducible R package and Shiny app for forward and backward citation chasing.
In ARISMA, snowballing is not a merely presentational addition, as it is a controlled enrichment step that should occur after the initial included set stabilizes and again when the map or synthesis suggests missing clusters. AI can help prioritize the snowballed pool by similarity, topic tagging, or study-design detection, but it should not silently filter the snowball yield without the same validation discipline used in ordinary screening.

{\bf Calibration.} 
Calibration is the key bridge between human review method and AI-assisted review method. It should happen before screening, before extraction, before categorization, and again when model prompts or settings change. A proper calibration exercise includes a pilot subset, blinded or semi-blinded human decisions, AI output comparison, adjudication of disagreements, and a protocolized revision of criteria or prompts. The living systematic review prompt-development study is particularly relevant \cite{homiar2025development} because it shows that strong performance did not come from a single prompt typed once into a chatbot; it came from structured development, testing, refinement, and consistency checks. ARISMA therefore treats calibration as a recurring methodological stage rather than a one-off preliminary check. 
Since hosted models can change without notice and corpora drift over time, ARISMA specifies {\em event-triggered} recalibration rather than a fixed universal interval. Recalibration on the sentinel and pilot sets is required whenever {\em i)} the model, version, interface, or decoding settings change; {\em ii)} prompts, eligibility criteria, or the codebook are amended; {\em iii)} corpus composition shifts materially, for example when a new database is added, a living-review update cycle begins, or retrieval volume grows beyond a pre-declared proportion; or {\em iv)} monitoring detects sentinel misses, unstable outputs, or a rising human-override rate. In living reviews, a small sentinel benchmark should additionally be re-run at every scheduled update cycle and logged in the validation matrix. Fixed time-based intervals may supplement, but never replace, these trigger conditions.

{\bf Language and coverage audit.}
AI assistance can skew evidence visibility toward English-language and terminologically mainstream literature: LLM screening performance varies across languages \cite{khraisha2024can}, and model-expanded queries may favor dominant vocabularies. ARISMA encourages research teams to report the language distribution of records at identification and after AI-influenced filtering steps, together with any language restrictions or translation workflows and their likely effect on coverage. Where non-English studies are eligible and resources allow, a fuller audit is a useful option: including non-English records in the sentinel and pilot sets, and reporting screening sensitivity stratified by language. If a marked asymmetry appears between the pre-AI and post-AI language distributions, it is worth investigating and documenting, and may warrant revised prompts, criteria, or human rescreening.

{\bf Data extraction.} 
Extraction is where many teams overestimate LLM maturity. LLMs are often good at pulling salient textual descriptions, eligibility rationales, intervention names, or author-reported conclusions. They are much less reliable when the target is a numerically exact datum embedded in prose, tables, figure labels, or complex study designs. A rapid feasibility study of GPT-4 for systematic-review extraction reported around 80\% overall accuracy, with better performance in some domains than others and specific difficulty with causal-inference methods and study design \cite{schmidt2024exploring}. More recent evaluation of AI-assisted extraction from randomized clinical trials showed a sharper contrast: for binary outcomes, some models achieved high accuracy for group sizes, but event counts were only moderate, and continuous-data elements such as means and standard deviations were poor, with accuracies as low as 24\% to 56\% depending on the model and variable \cite{yisha2026assessing}. A more favorable evaluation reached a compatible conclusion: benchmarked against a reference standard set by two independent human extractors, ChatGPT-4o showed high validity and reproducibility and was judged adequate as a {\em second} rater alongside a human reviewer rather than as a sole extractor \cite{motzfeldt2025chatgpt}. Taken together, these results strongly support a conservative rule: {\em AI may prepopulate extraction forms, but every numeric datum used in aggregation or synthesis must be checked against the source by a human reviewer}.

ARISMA extraction workflow separates fields into three classes. {\em Descriptive} fields such as country, study design, population label, or intervention family may be pre-extracted by AI and then quickly verified. {\em Interpretive} fields such as mechanism, implementation barrier, or thematic category may be AI-suggested but require human coding and reconciliation. {\em Critical quantitative} fields such as sample sizes, event counts, effect estimates, confidence intervals, follow-up times, and risk-of-bias support text require line-by-line or page-level human confirmation. The more the extracted datum affects a pooled estimate or a certainty judgment, the less acceptable the unverified AI extraction becomes.

{\bf Data aggregation and result gathering.} 
ARISMA distinguishes between {\em result gathering} and {\em data aggregation} because teams often merge them prematurely. Result gathering means assembling the verified evidence objects: included studies, linked reports, extraction forms, appraisal sheets, and justifications. Data aggregation means transforming those verified objects into evidence tables, grouped datasets, map cells, or synthesis-ready frames. AI can help normalize terminology, harmonize author labels, identify likely multiple-report clusters, or draft study profile cards, but aggregation must remain tied to a provenance-aware data structure. If a synthesized statement cannot be traced back to a verified extraction field, it does not belong in the review. This is especially important in LLM-assisted workflows because generative prose can sound definitive even when it is not grounded in the extracted dataset.

%\cite{campbell2020synthesis}

{\bf Categorization and keywording.} 
Mapping studies have long emphasized keywording and classification scheme construction as central steps; for example, keywording of abstracts is described as a way to iteratively build the categories that will structure the map \cite{petersen2008systematic}. LLMs can make this step much faster by proposing candidate tags, facet names, cluster labels, and latent-topic groupings from titles and abstracts. But those proposals are never the classification scheme itself. Under ARISMA, keywording is exploratory; categorization is confirmatory. The model may suggest the first, but the team must finalize the second through human consolidation, codebook definition, and back-application checks. A useful operational pattern is to let the model generate candidate keywords from a pilot set, then have two human reviewers collapse those into a controlled codebook and test inter-rater consistency before coding the full corpus.

{\bf Horizontal and vertical analysis.} 
These steps are especially useful in scoping, mapping, and evidence-gap work. ARISMA defines {\em horizontal analysis} as the cross-corpus view: distributions across years, geographies, study designs, data sources, topics, interventions, outcomes, or populations. It defines {\em vertical analysis} as the within-cluster view: a deeper examination of what a particular category or theme actually contains, how methods differ inside it, and what findings or tensions characterize it. AI is valuable in both layers, but in different ways. For horizontal analysis, AI can generate candidate facet combinations and visual summaries from structured data. For vertical analysis, AI can assist with within-cluster thematic condensation, especially if the source material has already been coded and verified. What it must not do is invent cross-tab counts, infer absent categories, or narrate causal conclusions from map frequencies alone.

{\bf Synthesis.}
Quantitative synthesis and qualitative or narrative synthesis need separate AI rules. For quantitative synthesis, AI may help identify missing cells, check arithmetic consistency, suggest subgroup structures, or draft ordinary-language explanations of already computed results, but statistical model choice, heterogeneity interpretation, certainty assessment, and final effect interpretation remain expert obligations. For non-meta-analytic quantitative synthesis, SWiM is directly relevant because it exists to reduce the opacity of narrative or alternative synthesis methods \cite{campbell2020synthesis}. For scoping and mapping reviews, synthesis often entails structured descriptions, thematic consolidation, and statements of implication rather than pooled estimates. In these settings, AI may help draft theme summaries or map narratives, but only from completed evidence tables and verified coding. Because LLMs can overgeneralize scientific findings, narrative synthesis is one of the places where unverified AI prose is most dangerous.

{\bf Methodological appraisal and quality assessment.}
ARISMA recommends an important distinction. Appraisal of review reports using validated checklists appears increasingly tractable for fine-tuned LLM support: a recent study found that a fine-tuned GPT-3.5 model achieved a mean accuracy of 96.5\% and a mean kappa of 0.90 when supporting methodological-quality assessment of systematic reviews using a validated 27-item tool \cite{marques2025use}. But appraisal of primary studies, especially nuanced risk-of-bias judgments, remains harder and should stay in a human-led workflow. 
%RobotReviewer is a useful precedent for automation in primary-study bias assessment because it extracts support text and risk judgments from trial reports, yet even that literature frames automation as support rather than as a replacement for methodological judgment.
In ARISMA, AI may therefore be used to pre-annotate appraisal forms and retrieve supporting text, but final risk-of-bias or certainty judgments must be reviewer-approved.

{\bf Expert validation of process and results.}
Consultation and expert review should be built into AI-assisted syntheses well before final proofreading. In the scoping review tradition, the literature argues that consultation can strengthen both process and product \cite{levac2010scoping}. In ARISMA, expert validation has four legitimate targets: the framing and scope, the search coverage and missing-study risk, the categorization or codebook, and the plausibility of the synthesis claims. The process should be documented. Domain experts should not merely {\em react} to the final paper; they should be asked specific questions such as whether sentinel studies are missing, whether category labels distort field practice, whether the map overstates maturity, and whether AI-drafted text collapses meaningful distinctions.

{\bf Threat logging and amendment handling during conduct.}
AI-supported reviews need a living error ledger. Every important model failure should be recorded: missed sentinel paper, unstable classification, hallucinated extraction, over-broad summary, or unexplained output shift after a model update. Amendments to prompts, tools, or validation thresholds should be dated and justified. This is not bureaucracy for its own sake. It is the minimum scaffolding required to make the review reproducible enough for peer evaluation. A review team that cannot reconstruct how AI affected the workflow has not really conducted an auditable evidence synthesis.

\section{Governance, Evidence, and Provenance}

%This section details how ARISMA moves from workflow description to governance. Sections 1 and 2 establish that AI can support multiple review tasks, but only under bounded and validated conditions. The purpose of this section is to define those conditions in operational terms. %Figures 4–6 collectively show how ARISMA governs AI use across time: Figure 4 specifies the validation cycle before and during deployment; Figure 5 shows how record flow, human checkpoints, and provenance should be integrated into an AI-aware evidence-flow architecture; and Figure 6 defines the verified chain from included reports to synthesis claims. Together, these figures articulate the paper’s central stance that methodological legitimacy depends not on using AI, but on controlling and documenting its influence. 

\subsection{Governance and validation logic}

%This subsection explains
%protocol-first governance,
%admissibility of AI,
%validation,
%calibration,
%override,
%logging,
%amendments, and
%iterative reassessment.
%**** MORE opening text before going to the Figure explanation***

The introduction of AI into evidence synthesis creates a methodological problem that classical review guidelines did not need to address explicitly: how to govern probabilistic systems whose outputs may vary across prompts, versions, contexts, and deployment settings. Conventional review methodology assumes that methodological steps are deterministic and reviewer-controlled. AI-assisted workflows challenge that assumption because model outputs may shift over time, may behave differently across corpora, and may appear persuasive even when unsupported by the source evidence. ARISMA therefore treats AI use not as a software convenience, but as a governed methodological intervention requiring predefined validation rules, auditability, explicit accountability, and continuous oversight.

\begin{figure}
    \centering
    \includegraphics[width=\linewidth]{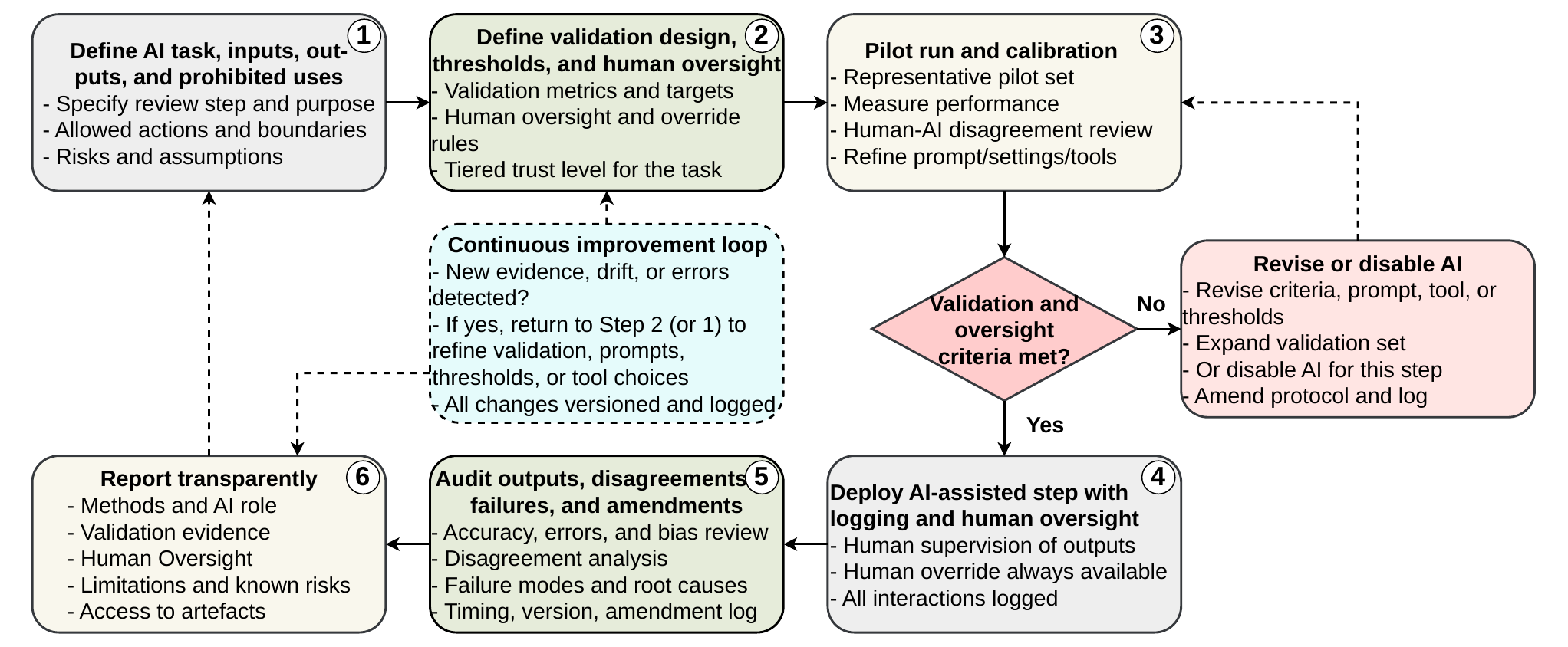}
    \caption{AI governance and validation cycle in ARISMA: protocol first, validation deployment, and continuous oversight. Evidence synthesists remain responsible for the review. AI is an inspected, benchmarked, logged, and reversible assistant, not an autonomous reviewer.}
    \label{gov}
\end{figure}

To operationalize this principle, ARISMA adopts a protocol-first governance logic. AI is not introduced into a review merely because a tool is available or performant in another study. Instead, the review team must first define the intended AI role, the boundaries of acceptable behavior, the validation criteria, the conditions under which the model may influence review decisions, and the situations in which AI support must be revised, restricted, or disabled entirely. Figure \ref{gov} summarizes this governance cycle and illustrates how validation, deployment, auditing, and reassessment are integrated throughout the lifecycle of an AI-assisted review.

Figure \ref{gov}  operationalizes ARISMA’s protocol-first governance logic. It shows that AI use is never justified by availability alone: the task, its boundaries, the validation criteria, human-override rules, and stopping conditions must be defined before deployment. The figure also makes reassessment explicit, because model suitability can change with corpus drift, prompt changes, interface updates, or newly observed failure modes.

%Figure \ref{gov} presents ARISMA’s governing cycle in six steps: define the task and its boundaries; define validation and oversight rules; run a pilot and calibrate; deploy under logging and human supervision; audit outputs, disagreements, failures, and amendments; and report the entire process transparently. Experts warned against assuming that models remain suitable over time. Therefore, the figure became cycle as model suitability is not a one-time property. Inputs change, prompts drift, source material varies, and hosted models may be updated outside the review team’s control. A model that was acceptable during pilot screening may become unacceptable after criteria changes, corpus expansion, or interface updates. Figure \ref{gov} therefore makes continuous reassessment a first-class methodological requirement rather than an optional quality-improvement step.

\subsection{AI-aware evidence flow and provenance}

Classical evidence-flow diagrams primarily document how records move through identification, screening, eligibility assessment, and inclusion. However, once AI systems participate in retrieval, screening, extraction, or synthesis, flow reporting alone becomes insufficient. Readers must additionally understand where AI entered the workflow, how AI outputs were validated, which decisions remained human-controlled, and how provenance was preserved from source documents to synthesis claims. In AI-assisted reviews, transparency therefore requires not only numerical accounting of records, but also methodological accounting of machine involvement.

\begin{figure}
    \centering
    \includegraphics[width=\linewidth]{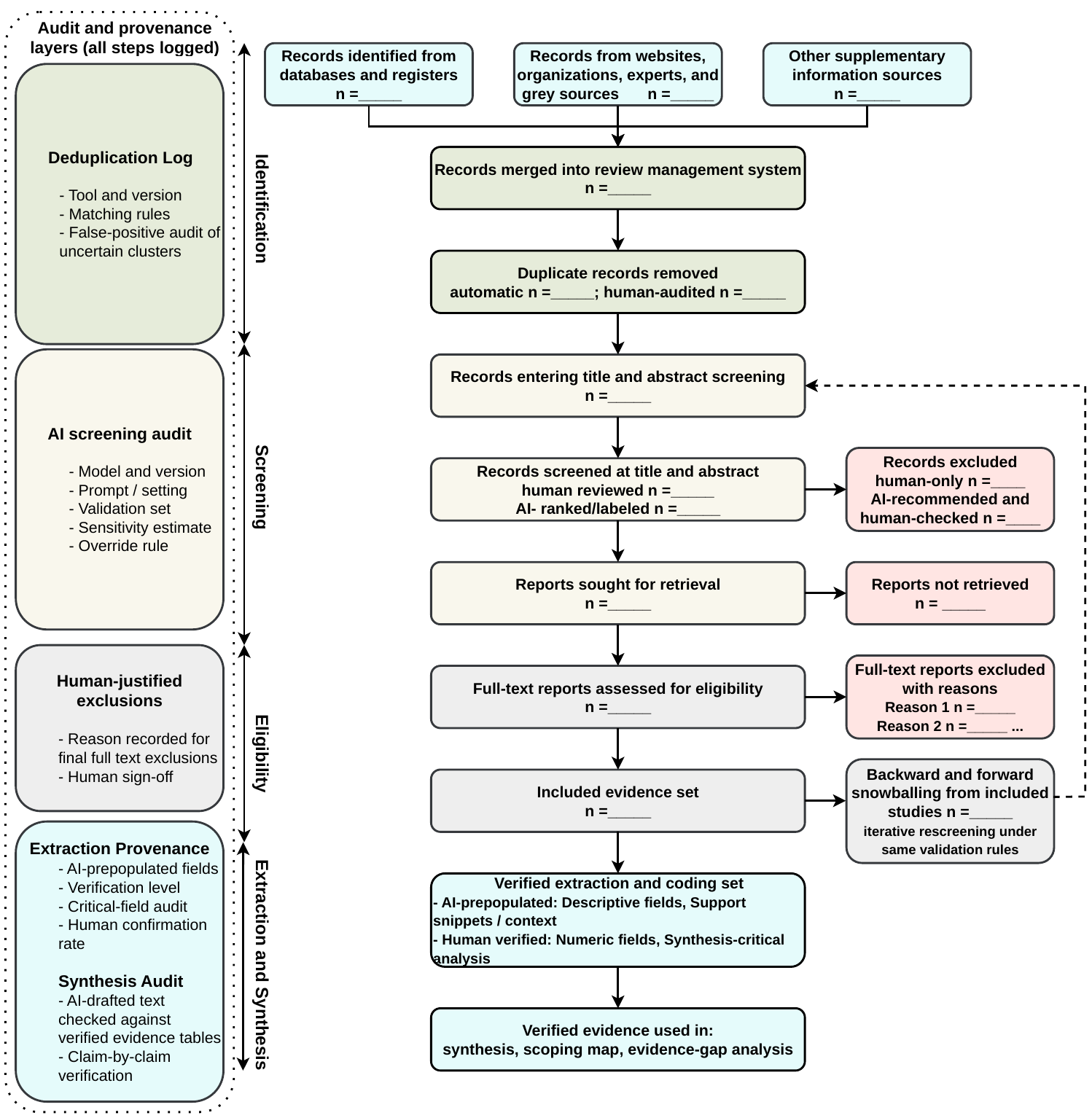}
    \caption{ARISMA evidence flow with iterative snowballing, provenance tracking, validation, and human checkpoints. n is the number of records/reports at that step. All AI outputs are inspected, validated, and overridden by humans as needed.}
    \label{flow}
\end{figure}

ARISMA extends traditional PRISMA-style evidence flow by embedding provenance layers, validation checkpoints, human override rules, and audit traces directly into the evidence pipeline. The objective is to demonstrate how scientific control was maintained throughout the review rather than showing how many records were processed. Figure \ref{flow} operationalizes this AI-aware evidence-flow logic and demonstrates how auditability, traceability, and iterative enrichment are integrated into the review lifecycle.

%Figure \ref{flow} extends conventional PRISMA-style flow reporting into an AI-aware evidence-flow model. Once AI contributes to retrieval, screening, extraction, or synthesis, record counts alone are no longer sufficient; readers must also see where AI entered the workflow, what validation rules applied, and where human sign-off remained mandatory. The figure therefore combines evidence flow with provenance, audit, and override layers.

%Figure \ref{flow} shows once AI touches identification, screening, extraction, or synthesis, purely numeric flow reporting is no longer sufficient. The review must also make visible which stages were AI-assisted, what validation and override rules applied, and how provenance was maintained from records to synthesis-ready evidence. That is why audit layers are integrated for deduplication, AI-supported screening, extraction provenance, and synthesis verification. %These layers convert a record-flow diagram into an accountability diagram. 

%The most important textual clarification to add here is that 
Provenance is cumulative. Deduplication provenance records how overlapping source exports were resolved. Screening provenance records what the model recommended, what humans overrode, and how false exclusions were prevented. Extraction provenance records which fields were AI-prepopulated, which were fully human-entered, and which underwent source-level confirmation. Synthesis provenance records whether interpretive prose was derived from verified tables or generated more freely. 
%Experts also called for {\em “a chain of custody showing how many records and which ones were influenced by AI”}. 
%The figure therefore shows a layered chain of evidence custody. Without that chain, readers cannot tell whether the review conclusions are anchored in reviewed evidence or have been reshaped by opaque tool behavior.
Consultation feedback also called for a clearer chain of evidence custody showing where AI influenced records and outputs. Figure 5 therefore presents a layered provenance structure. Without that structure, readers cannot determine whether the review conclusions are anchored in reviewed evidence or have been reshaped by opaque tool behavior.

%One of the central risks of LLM-assisted evidence synthesis is that fluent outputs can obscure weak provenance. A synthesis statement may appear coherent and authoritative even when the underlying extraction chain is incomplete, partially verified, or disconnected from the original source material. ARISMA therefore treats provenance as a continuous methodological chain linking every synthesis claim back to verified source evidence, documented coding decisions, and human-confirmed extraction outputs.

%This requirement becomes particularly important in AI-assisted workflows because models can prepopulate extraction forms, propose categories, generate thematic summaries, and draft narrative text at multiple stages of the review. Without a structured verification chain, errors introduced early in the process may propagate silently into synthesis and interpretation. Figure \ref{prov} illustrates the verified provenance chain proposed by ARISMA, showing how evidence moves from included studies through extraction, verification, categorization, and synthesis while remaining continuously auditable and human-accountable.

%*** Description of provenance with this figure is missing too ***

\begin{figure}
    \centering
    \includegraphics[width=\linewidth]{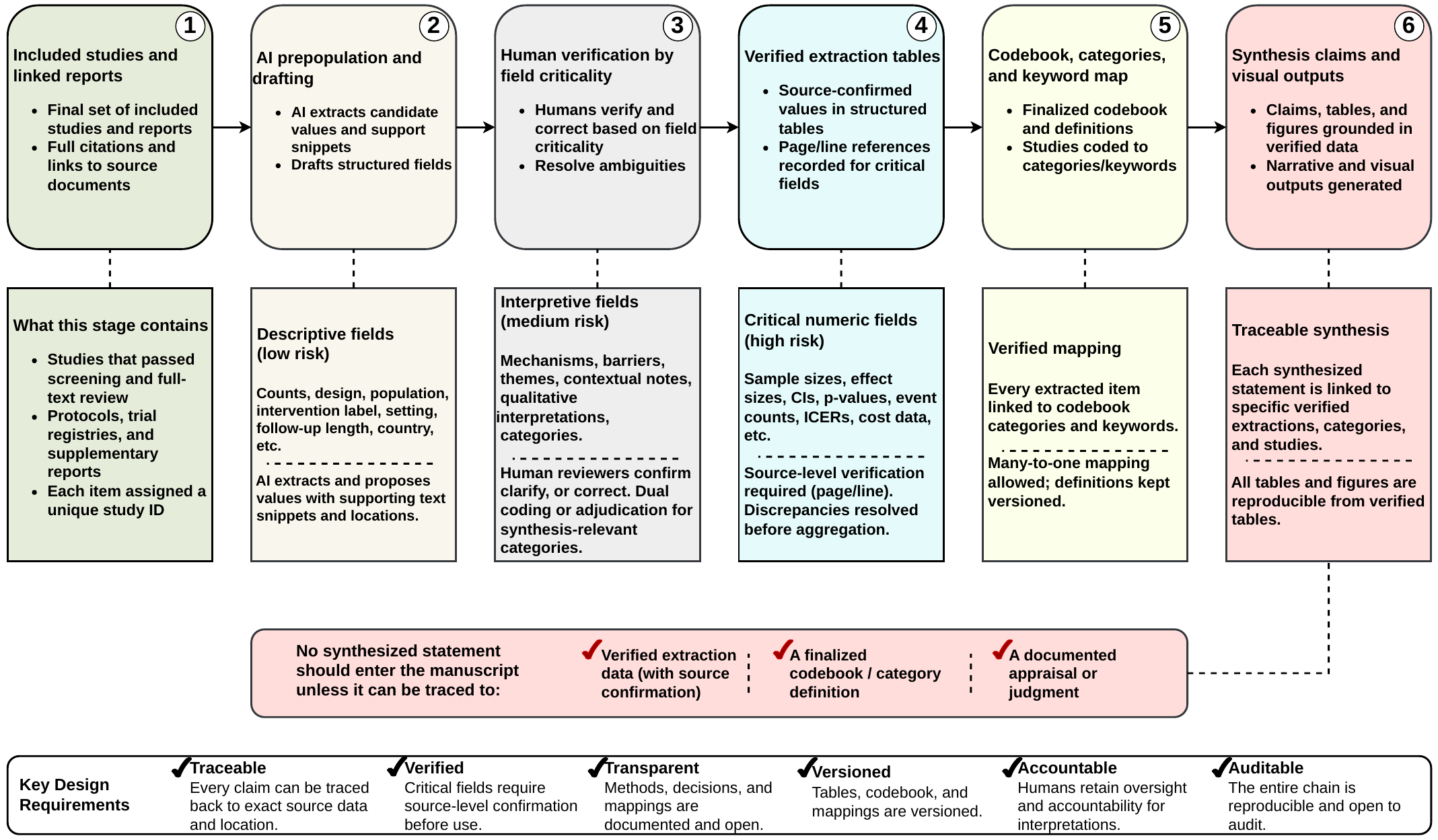}
    \caption{Verified provenance chain from extraction to synthesis. AI can prepopulate
and draft, but verified evidence tables remain the source of truth.}
    \label{prov}
\end{figure}

%Figure \ref{prov} presents the verified provenance chain from included studies to synthesis claims. Its purpose is to show that AI may accelerate movement through the review pipeline, but it cannot replace the source-confirmed extraction layer as the source of truth. The figure also makes field criticality explicit by separating lower-risk descriptive fields, medium-risk interpretive fields, and high-risk quantitative fields that require source-level human verification.

%Figure \ref{prov} is a verified evidence chain from source documents to manuscript claims. Its sequential stages, i.e., included studies, AI prepopulation, human verification by field criticality, verified extraction tables, codebook/category assignment, and synthesis claims/visual outputs, show that AI assistance is acceptable only when it remains downstream of source material and upstream of human confirmation. In other words, AI can accelerate movement through the chain, but it cannot terminate the chain. The source of truth remains the verified extraction layer, not the model output layer. 

Consultation feedback supported triaging data fields according to their consequences for later synthesis. 
%Experts insisted on triaging data fields by risk.
As Figure \ref{prov} shows, descriptive fields such as study setting, country, or intervention family are lower-risk and can plausibly be prepopulated by AI, subject to verification. Interpretive fields such as themes, mechanisms, barriers, or contextual labels are medium-risk because they shape later categorization and narrative synthesis; here AI suggestions may be useful, but human coders must consolidate and adjudicate. Critical quantitative fields, i.e., sample sizes, effect estimates, event counts, confidence intervals, follow-up durations, cost values, or any value that directly affects aggregation, are high-risk and should never enter synthesis without source-level human confirmation.  
%An expert observed that {\em “the figure now draws a bright line: AI can help propose values and snippets, but numbers and themes that feed into synthesis must be verified by human reviewers.”} 
This distinction establishes a clear boundary: AI may propose values and supporting snippets, but quantitative values and interpretive categories that feed into synthesis require human verification. 
This triage is also strongly supported by previously discussed literature, showing that extraction performance is uneven across field types and substantially worse for some numeric and inferential variables than for simpler descriptive content. 
No synthesized sentence should appear in the manuscript unless it can be traced back to a verified field, a codebook decision, or a source-confirmed analytic output. %This is exactly where ARISMA becomes more than an AI checklist. It becomes a provenance discipline. That framing is likely to resonate strongly with a journal concerned with AI accountability and scholarly integrity. 

\subsection{Legal, Privacy, and Infrastructure Governance}

AI-assisted review workflows raise legal and privacy questions that classical review guidelines did not need to address explicitly. In a conventional review, bibliographic records, PDFs, extraction tables, and notes are typically processed within citation managers, spreadsheet environments, and team-shared repositories. In an LLM-assisted workflow, those same materials may be transmitted to third-party services, stored in vendor infrastructure, retained in logs, or processed across jurisdictions. The methodological consequence is that a review team can no longer treat {\em using a model} as a purely technical choice. It is also a data-governance choice. For ARISMA, the default rule should therefore be to classify review inputs by sensitivity before any AI use begins. Public metadata such as titles, abstracts, keywords, and database-exported bibliographic records are usually the lowest-risk inputs. Full texts, unpublished manuscripts, reviewer comments, extraction sheets containing copied passages, interview transcripts, and any documents containing personal or confidential data are higher-risk inputs and should be governed accordingly. 

%There are three following infrastructure classes to be considered.
Three infrastructure classes should be considered:

\begin{itemize}
    \item Public consumer interfaces, which are the least suitable route for high-stakes review work because they often provide the weakest contractual guarantees, the least predictable retention behavior from the user’s perspective, and limited auditability. 
    \item Enterprise or institutionally governed hosted interfaces, which may provide contractual, access-control, logging, or retention safeguards but still require explicit review-team documentation of what was uploaded, for what purpose, under what access mode, and with what institutional approval.
    \item Local or controlled institutional deployment, including on-premise or research-cluster execution, which is often preferable for sensitive documents because it reduces exposure to external processing and gives the team greater control over versioning, logging, and retention. %ARISMA need not prescribe one infrastructure universally, but it should require authors to disclose which class they used for each AI-supported task. 

\end{itemize}

Privacy assurance should be addressed around governance principles. Teams should at least articulate lawfulness, transparency, data minimization, access control, storage limitation, and integrity/confidentiality safeguards for AI-assisted processing of review materials.  

Copyright and manuscript confidentiality should also be considered. Review teams often work with publisher PDFs, subscription content, or institutionally licensed materials. ARISMA recommends that authors do not upload full texts, unpublished manuscripts, or any sensitive peer-review material to external generative-AI services unless they have determined that such processing is legally and contractually permissible and consistent with institutional policy.

\section{Tools and Trends}

\subsection{Tool Support}

As detailed above, systematic reviews involve a series of labour-intensive tasks: defining scope, formulating search strategies, retrieving and deduplicating citations, screening titles and abstracts, chasing references, extracting data, assessing risk of bias, synthesizing findings and reporting. Over the last decade, dozens of tools using rule-based, machine-learning and natural language processing techniques have been developed to automate or semi-automate these tasks. ARISMA does not endorse any specific tool; rather, it classifies tools into four categories and provides governance principles for their use.

\begin{enumerate}
    \item {{\bf \em Search strategy design and translation:}
    \begin{itemize}
        \item Visual query builders and synonym finders (e.g. 2dSearch\footnote{\url{https://www.2dsearch.com/}}) help reviewers construct complex search strings by dragging and connecting concepts on a canvas. These interfaces improve transparency and reduce syntactic errors.
        \item Search-translation utilities (e.g. Polyglot Search Translator, part of TERA \footnote{\url{https://tera-tools.com/help/polyglot}}) convert a search string from one database syntax to another and often integrate controlled vocabularies. 
        Search-translation utilities such as the Polyglot Search Translator can substantially reduce the manual effort required to adapt search strategies across databases while helping preserve structural consistency and recall when compared with fully manual translation workflows.
        %Randomised trials show that they reduce translation time by around 10 minutes per database without sacrificing recall.
        \item Term-expansion tools (such as PubReMiner\footnote{\url{https://hgserver2.amc.nl/cgi-bin/miner/miner2.cgi}}) suggest synonyms and spelling variants; they should be used as drafting aids and not as a substitute for librarians’ expertise.
    \end{itemize}
    }
    \item {{\bf \em Citation retrieval, deduplication and snowballing:}
    \begin{itemize}
        \item Automated deduplicators such as ASySD\footnote{\url{https://camaradesuk.github.io/ASySD/}} and the TERA\footnote{\url{https://tera-tools.com/}} compare fields (title, author, DOI, year) using rule-based matching and ML. %evaluations report sensitivities above 0.90, greatly outperforming manual and EndNote-only deduplication. 
        Evaluations of ASySD reported high sensitivity and specificity for duplicate detection across multiple systematic-review datasets, outperforming EndNote-based automatic deduplication approaches while substantially reducing manual review burden \cite{hair2023automated}. However, ambiguous pairs still require human checking. ARISMA therefore designates deduplication as an assistive task, with logs of removed pairs retained for audit.
        \item Citation-chasing tools (e.g. Citationchaser\footnote{\url{https://estech.shinyapps.io/citationchaser/}}, Paperfetcher\footnote{\url{https://paperfetcher.github.io/}}, SpiderCite\footnote{\url{https://tera-tools.com/help/spidercite}}) can accelerate forward and backward snowballing. Validation studies show that co-citation ranking can retrieve most eligible articles in the top ranks, saving time when more than 500 titles need to be screened \cite{janssens2020novel}. Nevertheless, users must verify that the seed set is comprehensive and that relevant but infrequently cited studies are not missed. ARISMA recommends using citation-chasing tools only after the primary search has been benchmarked against sentinel studies.
    \end{itemize}}
    
    \item {{\bf \em Screening and prioritisation:}
    \begin{itemize}
    \item Active-learning screeners such as Rayyan\footnote{\url{https://www.rayyan.ai/}}, Abstrackr\footnote{\url{https://abstrackr.com/}}, Research Screener\footnote{\url{https://researchscreener.com/}}, ASReview\footnote{\url{https://asreview.nl/}}, SWIFT-Review \footnote{\url{https://www.sciome.com/swift-review/}}, SWIFT-Active Screener\footnote{\url{https://www.sciome.com/swift-activescreener/}}, RobotAnalyst\footnote{\url{https://www.nactem.ac.uk/robotanalyst/}} and commercial platforms like DistillerSR\footnote{\url{https://www.distillersr.com/}} and Covidence\footnote{\url{https://www.covidence.org/}} rank or classify records based on labels provided during screening. %User studies consistently report 40-90\% reductions in workload with minimal loss of recall, but they also highlight the risk of user complacency and false negatives. 
    Recent evaluations of AI-assisted screening pipelines reported Work Saved over Sampling at 95\% recall values ranging from approximately 49\% to 87\% under benchmark conditions \cite{kataoka2026tiab}.
    These evaluations also emphasize the importance of rigorous calibration, stopping criteria, and continued human verification to minimize the risk of false-negative exclusions. 
    %ARISMA therefore treats screening tools as adjudicative assistants: models may rank or suggest decisions, but human reviewers remain final arbiters. 
    Validation must include a pilot set, a target sensitivity (e.g. 95\% recall), an explicit stopping rule (as provided by recall estimators in SWIFT-Active Screener), and documentation of all human overrides.
    \item Platforms that combine screening with extraction and synthesis (e.g. Nested Knowledge\footnote{\url{https://nested-knowledge.com/}}, RevMan Web\footnote{\url{https://revman.cochrane.org/info}}, JBI SUMARI\footnote{\url{https://sumari.jbi.global/}}) enforce structured workflows and data formats but are not yet AI-driven. They can be used as organizational scaffolds around which AI-assisted tasks are inserted.
    \end{itemize}}
    \item {{\bf \em Data extraction and risk-of-bias assessment:}
    \begin{itemize}\item %Tools like ExaCT and RobotReviewer automatically identify sentences describing trial characteristics and generate risk-of-bias judgments. 
    Recent extraction and appraisal systems (e.g., AIDE\footnote{\url{https://github.com/noah-schroeder/AIDE}}) 
    can pre-populate structured fields and surface support snippets, but current studies show mixed, item-dependent performance. They are most defensible as assistive tools for descriptive fields and reviewer workflow support; synthesis-critical values and final appraisal judgments still require explicit human verification \cite{van2023artificial,burns2024using}. 
    %can support identification of study characteristics, outcome descriptions, and methodological information from full-text articles. However, current evidence supports their use primarily as assistive technologies requiring explicit human verification before extracted data or risk-of-bias judgments are incorporated into synthesis.
    %Evaluations report recalls around 0.90 and accuracies approaching human performance, but outputs still require verification \cite{van2023artificial, burns2024using}.
    %In constrained biomedical abstract-classification settings, supervised ML classifiers achieved approximately 90\% classification accuracy and reported potential workload reductions of around 70\%.
    %ARISMA classifies extraction and appraisal as adjudicative tasks: AI may prepopulate fields and highlight relevant sentences, but human reviewers must confirm and correct entries.
    \item End-to-end platforms (e.g. Covidence, DistillerSR) allow customized extraction forms but do not fully automate extraction. As such, they serve as management environments rather than AI tools.
    \end{itemize}}
\end{enumerate}

Across these tool categories, ARISMA applies the same governance rule: automation may accelerate bounded tasks, but methodological accountability remains with the review team. The required level of validation and human verification increases with the epistemic consequence of the task. Three principles apply:

\begin{itemize}
    \item {\bf \em Benchmark and calibrate:} Before any AI component is used in production, run a pilot on a labeled subset. Evaluate sensitivity (recall) and precision (work saved) and set thresholds. If the model fails to meet the threshold, adjust the prompt or algorithm or switch to a different tool. Calibration results should be reported in the methods section and retained for audit.
    \item {\bf \em Document and audit:} All AI-driven actions must be logged, i.e., the model name and version, the date of use, key settings, the labeled pilot set, and the number of records affected. For deduplication, record pairs removed; for screening, record AI-proposed exclusions and human overrides; for extraction, record AI-filled fields and human edits. These logs enable reproducibility and support peer review.
    \item {\bf \em Assign human sign-off:} AI can prioritize work but may not make final judgments on eligibility, risk of bias, or synthesis. For high-consequence tasks, at least one experienced reviewer must verify outputs. Multi-model agreement (running two independent models and comparing results) can be used to increase confidence, but disagreement must always be resolved by humans.
\end{itemize}

By following these principles, researchers can take advantage of the efficiency gains of automation while preserving scientific accountability.

\subsection{Sustainability and Efficiency}

AI deployment depends on electricity-intensive data-center or accelerator infrastructure, and international energy analyses now treat AI as a meaningful driver of future data-center electricity demand. At the same time, inference costs are highly variable across models, hardware stacks, workloads, and serving configurations, and careful optimization can substantially reduce energy use relative to unoptimized inference pipelines. The practical implication for ARISMA is not that teams must compute exact carbon footprints for every review. Rather, teams should prefer the least resource-intensive configuration that still meets methodological requirements. In practice, that means favoring bounded task-specific tools over open-ended generation, smaller or local models for routine classification when performance is adequate, batch processing over repeated interactive prompting, and human escalation only for ambiguous or high-consequence cases. 

Multi-agent debate, large reasoning models, repeated self-consistency sampling, and excessively long generative prompts can increase inference cost without proportionate methodological gain. ARISMA thus discourages {\em compute escalation by default}. More compute is justified only when it measurably improves validation outcomes or reduces human burden without increasing epistemic risk. 
So that efficiency claims can be reported rather than merely asserted, ARISMA recommends a minimal {\em resource-disclosure set} for every AI-assisted step: the model class, size, and version; the number of calls and records processed; total input and output tokens for API-based use; and measured energy in kWh where the team controls the hardware, for example via software-based meters. 
From these primitives, teams may report derived intensity ratios such as tokens per screened record or {\em tokens per validated record}, provided these are labeled as workload proxies rather than energy measurements. %, since token counts do not translate into energy or carbon across models, hardware stacks, and serving configurations, and cross-review comparisons are further confounded by corpus size and prevalence.
Reporting at this level is already feasible: recent screening-tool evaluations log per-record token consumption and API cost \cite{kataoka2026tiab}, and energy analyses of AI inference provide the reference points needed to interpret such disclosures \cite{iea2025energyai,luccioni2024power}.

%From these primitives, teams may report derived intensity ratios such as tokens per screened record or {\em tokens per validated record}, provided these are labeled as workload proxies rather than energy measurements, since token counts do not translate into energy or carbon across models, hardware stacks, and serving configurations, and cross-review comparisons are further confounded by corpus size and prevalence. Reporting at this level is already feasible: recent screening-tool evaluations log per-record token consumption and API cost \cite{kataoka2026tiab}, and energy analyses of AI inference provide the reference points needed to interpret such disclosures \cite{iea2025energyai, luccioni2024power}.

%******TO ADD******* THE TABLE SHOULD HAVE SOME SHORT DESCRIPTION*** The section is too short **** There was another document with explnatation find it

%%
\section{Reporting and Validation Checklist}

%PRISMA 2020 remains the right reporting anchor for systematic reviews, PRISMA-S remains the right anchor for search transparency, PRISMA-ScR remains the right extension for scoping reviews, PRISMA-P remains central for protocols, and SWiM remains essential for non-meta-analytic quantitative synthesis. ARISMA does not replace these standards. Instead, it adds an AI layer to them. In other words, a review should still satisfy the classical reporting items for its design, but whenever AI is used it should also disclose what system was used, for what task, on what inputs, under what settings, with what validation evidence, under what human oversight rule, and with what remaining limitations. 

Tables \ref{items} and \ref{matrix} convert ARISMA from a general framework into a submission-ready reporting and validation instrument. Table \ref{items} specifies what AI use must be disclosed; Table \ref{matrix} specifies what must be validated before AI output can influence the review.

%ARISMA adds an AI-governance layer to previous systematic review guidelines. Table \ref{items} extends classical reporting expectations by specifying what must additionally be disclosed when AI shapes any part of the review lifecycle. Table \ref{matrix} then operationalizes those expectations as validation obligations: it states what the team must have tested, what minimum evidence of adequacy should exist, and where human approval remains mandatory. The two tables transform AI use from a vague methods note into an inspectable methodological object.

% Preamble
%
%\vspace{-2em}
\subsection{ARISMA Item Checklist}

%The ARISMA Item Checklist (Table \ref{items}) does more than extend PRISMA; it reframes reporting as accountability. Beyond listing tools or models used, authors must specify how each AI component was tested and controlled. For example, under “Selection process” the checklist asks for the calibration set, the target recall threshold, whether the model acted as a ranker or a recommender, and the number of human overrides. Experts stressed that this level of detail makes AI use inspectable, turning the checklist into a pilot script rather than a punitive ledger. %Consequently, the final checklist cross-references the validation matrix and provides examples of acceptable disclosures.

%\vspace{-2em}
\newcolumntype{L}[1]{>{\raggedright\arraybackslash}p{#1}}

\begin{table}
\caption{Reporting items and additional AI/LLM reporting requirements}
\centering
\scriptsize
\setlength{\tabcolsep}{3pt}

\begin{tabular}{@{}L{0.16\textwidth}L{0.33\textwidth}L{0.50\textwidth}@{}}
\toprule

\textbf{Item} &
\textbf{What should be reported} &
\textbf{Additional AI and LLM reporting required} \\
\midrule

Title &
Identify the review type clearly &
State if the review was AI-assisted \vspace{.4em}\\ 

Abstract &
Structured summary of rationale, methods, results, and conclusions &
Name the AI-supported steps and the level of human oversight \vspace{.4em}\\

Rationale &
Why the review was needed &
Why AI was considered necessary or useful for this workflow \vspace{.4em}\\

Objectives &
Explicit review question and scope &
Which steps AI was allowed to influence and which were human-only \vspace{.4em}\\

Review design &
Systematic, scoping, mapping, EGM, rapid, or other &
Why the chosen AI role is appropriate for that review family \vspace{.4em}\\

Eligibility criteria &
Full inclusion and exclusion criteria &
How criteria were translated into prompts, rules, or labels \vspace{.4em}\\

Information sources &
Databases, registers, websites, grey sources, handsearching &
Whether AI assisted source discovery, query iteration, or web searching \vspace{.4em}\\

Search strategy &
Reproducible search strings and filters &
Prompt text or logic used to generate candidate terms; whether PRESS or equivalent review was done \vspace{.4em}\\

Record management &
Import, normalization, and deduplication methods &
Tool used for deduplication; human checks for ambiguous duplicates \vspace{.4em}\\

Selection process &
Who screened what, in how many stages, with what consensus process &
Validation set, sensitivity target, exclusion rule, override rule, whether AI acted as ranker, second screener, or exclusion recommender, and the language distribution of records before and after AI-influenced filtering \vspace{.4em}\\

Supplementary searching &
Snowballing, citation chasing, handsearching, expert contact &
Whether AI prioritized or tagged supplementary-search yields \vspace{.4em}\\

Data items &
What variables were extracted &
Which variables were AI-prepopulated and which were human-only \vspace{.4em}\\

Extraction process &
Form design, pilot testing, reviewer arrangement &
Model, version, prompt, format constraints, provenance capture, and human verification rate \vspace{.4em}\\

Appraisal or quality assessment &
Tool used and how judgments were reached &
Whether AI pre-annotated items or support text; who finalized judgments \vspace{.4em} \\

Synthesis methods &
Grouping, aggregation, analysis, visualization, and uncertainty handling &
Whether AI drafted themes, clusters, or narrative summaries, and how those drafts were checked \vspace{.4em}\\

Results of search and selection &
Counts and flow of records &
Whether AI changed the candidate pool and, if so, how validation was demonstrated \vspace{.4em}\\

Results of extraction and appraisal &
Study characteristics and quality findings &
Error rates or discrepancy rates between AI outputs and human-verified values \vspace{.4em}\\

Results of synthesis &
Main findings, heterogeneity, themes, gaps, or map structures &
Whether any claim in the synthesis originated from AI drafting and how it was linked to verified evidence tables \vspace{.4em}\\

Limitations &
Review limitations and evidence-base limits &
Known AI limitations, model instability, domain mismatch, unresolved error modes, and any observed differences in performance or coverage across languages \vspace{.4em}\\

Registration and protocol &
Registry, protocol access, amendments &
AI methods appendix location and all AI-related amendments \vspace{.4em}\\

Funding and conflicts &
Financial and non-financial disclosures &
Vendor relationships, paid subscriptions, API access, model licenses, and data-governance constraints \vspace{.4em}\\

Availability &
Data, code, extraction forms, appendices &
Prompts, model metadata, validation scripts, and audit logs insofar as legal and ethical constraints allow \\

\bottomrule
\end{tabular}
\label{items}
\end{table}

Table \ref{items} is a reporting instrument that prevents {\em AI-assisted} from becoming an empty label. In conventional methods sections, authors can sometimes mention automation in a single sentence, for example, by saying that a model assisted with screening or drafting. Under ARISMA, that is insufficient. Readers need to know what system was used, on what input, for which task, under which restrictions, with what validation evidence, and with what human override rule. The checklist therefore complements PRISMA-style reporting by making AI use legible at the same level of transparency expected for search strategies, eligibility criteria, screening procedures, or synthesis methods. 
%Experts stressed that this level of detail makes AI use inspectable, turning the checklist into a pilot script rather than a punitive ledger.
Consultation feedback indicated that this level of detail makes AI use more inspectable and allows the checklist to function as an operational reporting aid rather than only as a disclosure list.

%Table \ref{items} is the manuscript-facing expression of the governance logic developed earlier in the paper. 
% 
%The title and abstract should state whether the review was AI-assisted. The methods should document substantive LLM use because the journal explicitly requires this and does not permit LLM authorship. The Declarations section should include competing interests, data availability, ethics statements where relevant, and any constraints arising from proprietary or privacy-sensitive data. %For a double-anonymous submission, supplementary materials and repository links must also be handled in a way that preserves blinding. These are not peripheral editorial details; they are part of the paper’s accountability infrastructure.

%The checklist above is meant to be applied across a broad review families. In manuscript preparation, teams should crosswalk each ARISMA item against the full reporting guideline that governs their specific review type. 

%%
\subsection{Validation Matrix}

%The validation matrix is the operational heart of the guideline, because it prevents teams from treating {\em AI-assisted} as a vague productivity claim. 

The validation matrix translates ARISMA’s principles into step-specific evidentiary thresholds. %Each row identifies a risk (e.g. missed sentinel studies, false duplicate removal, extraction error), defines what must be validated, and specifies the evidence required.
%Experts emphasized that the burden of proof should rest on the AI: failing a validation test returns the process to calibration. The matrix therefore distinguishes between passing a threshold (a permission to proceed) and routine operation (which remains under human supervision).
%Experts emphasized that the burden of proof should rest on the AI: failing a validation test returns the process to calibration.
The consultation reinforced the principle that the burden of proof should rest on the AI-assisted step: failing a validation test returns the process to calibration. 
The matrix therefore distinguishes between passing a threshold (a permission to proceed) and routine operation (which remains under human supervision). 
Those permissions are also revocable rather than permanent, which is why the matrix closes with a recalibration row: a threshold passed under one model version, prompt, or corpus does not license continued use once any of them changes.

\begin{table}[htbp]
\caption{Validation requirements and minimum ARISMA evidence}
\centering
\scriptsize

\begin{tabular}{@{}L{0.26\linewidth}L{0.31\linewidth}L{0.37\linewidth}@{}}
\toprule

\textbf{Step} &
\textbf{What must be validated} &
\textbf{Minimum ARISMA evidence} \\
\midrule

Search-term generation &
Recall of known or sentinel studies &
Pilot benchmark set, documented failures, revised search after peer review \vspace{.4em} \\

Deduplication &
False positives and false negatives &
Tool log plus human audit of uncertain duplicate clusters \vspace{.4em} \\

Title and abstract screening &
Sensitivity for relevant records &
Human-coded pilot set, error review, explicit rule for handling AI exclusions \vspace{.4em} \\

Full-text screening &
Justification quality and false exclusions &
Human consensus sign-off on all final exclusions \vspace{.4em} \\

Extraction of descriptive fields &
Agreement with source papers &
Random or full audit, depending on field criticality \vspace{.4em} \\

Extraction of numeric fields &
Exactness of synthesis-critical values &
Full human verification of all values used in aggregation or meta-analysis \vspace{.4em} \\

Coding, keywording, categorization &
Stability of code assignment &
Human-reviewed codebook plus audit of assigned clusters \vspace{.4em} \\

Appraisal support &
Agreement with validated human judgments &
Item-level comparison and reviewer sign-off \vspace{.4em} \\

Narrative drafting &
Fidelity to verified evidence tables &
Sentence-level verification for all claims that affect conclusions \vspace{.4em} \\

%Final synthesis &
%No unsupported or AI-invented claims &
%Senior-methodologist review and domain-expert plausibility check \\
Final synthesis &
No unsupported or AI-invented claims &
Senior-methodologist review and domain-expert plausibility check \\\\

Recalibration (all AI-assisted steps) &
Continued validity after model, prompt, criteria, or corpus change &
Re-run sentinel benchmark on each trigger event; dated amendment log linking the trigger to the recalibration result \\

\bottomrule
\end{tabular}
\label{matrix}
\end{table}

Table \ref{matrix} functions to stop teams from describing an AI-supported review as rigorous when the underlying tool was never meaningfully tested for the role it played. Each row therefore answers three questions: what can go wrong at this step, what exactly must be validated before AI output is trusted, and what minimum audit trail should remain available after the review is completed. For search-term generation, the relevant concern is recall of known studies. For deduplication, it is the balance of false positives and false negatives. For screening, it is sensitivity for relevant records and the handling of unjustified exclusions. For extraction, it is agreement with the source and source-level confirmation of synthesis-critical values. 
%For drafting, it is fidelity to verified evidence tables and the absence of AI-invented claims.
For drafting, it is fidelity to verified evidence tables and the absence of AI-invented claims. For recalibration, it is whether a previously granted permission still holds after the model, prompt, criteria, or corpus has changed, evidenced by a re-run sentinel benchmark and a dated amendment log.

%For drafting, it is fidelity to verified evidence tables and the absence of AI-invented claims. %Table \ref{matrix} should be presented not as an optional appendix, but as a compact specification of review-team obligations. 

The matrix also helps reviewers and editors. Editors can use it to judge whether an AI-assisted paper remains under human scientific control. Peer reviewers can use it to determine whether the authors’ methodological claims are supported by validation artefacts rather than by confidence in a model brand. In other words, Table \ref{matrix} turns ARISMA into usable methodology at submission and peer-review time, not only during review conduct.

\section{Threats to Validity}

{\bf Design threat.}
The first threat is choosing the wrong type of review and then trying to compensate with technology. AI cannot rescue a poorly framed evidence-synthesis design. If a team really needs a focused causal answer, a vague scoping review with impressive clustering graphics is still the wrong product. Conversely, if a field is conceptually fragmented and immature, a false impression of precision produced by premature quantitative synthesis is also a design error. AI can intensify either mistake by making output appear more polished than it is.

{\bf Retrieval threat.}
Search incompleteness remains one of the most serious threats in all evidence syntheses. LLM-assisted query generation is not yet reliable enough to replace expert search design, and even strong systems can miss terminologically unusual but relevant studies. That is why ARISMA insists on benchmark sets, PRESS-style review, and complementary methods such as citation chasing. Snowballing itself brings another threat: citation databases differ in coverage, and forward-chasing yields are database-dependent.

{\bf Screening threat.}
Screening performance is context-sensitive. The encouraging GPT-4 and GPT-4o studies were conducted under constrained conditions, with refined prompts, particular prevalence structures, and explicit benchmarking. That does not mean their performance will transfer unchanged to a different domain, language mix, or eligibility regime. Automation complacency is therefore a major risk: teams may remember the success story but forget the validation conditions that made it possible. 
Language coverage is part of the same threat: if models screen or rank non-English records less reliably, the corpus can drift toward English-only evidence without the team noticing, which is why ARISMA encourages reporting the language distribution before and after AI-influenced filtering.

%Language coverage is part of the same threat: if models screen or rank non-English records less reliably, the corpus can drift toward English-only evidence unless the language and coverage audit described in Section 3.1 is enforced.

{\bf Extraction and appraisal threat.}
Extraction errors can be subtle and high impact. Numeric inaccuracies, misread tables, confusion between arms or time points, and conflation of reported and inferred values may all survive into synthesis if extraction is not fully checked. Appraisal support tools face a similar boundary problem: they may retrieve supporting text efficiently, but nuanced methodological judgments still require domain and design knowledge. The current literature supports assistance, not abdication.

{\bf Generative reasoning threat.}
The most distinctive LLM risk is not formatting error but epistemic drift. Models can overgeneralize, omit qualifiers, smooth over contradictions, and write a more coherent claim than the underlying evidence actually permits. Studies show polished scientific summaries can still overstate what the original papers justify. For review manuscripts, that means AI-generated discussion text is a threat surface, not a neutral convenience \cite{peters2025generalization}.
%\vspace{1em}

{\bf Reproducibility threat.}
Closed models change over time, prompting interfaces hide implementation details, and even low-temperature settings do not guarantee identical outputs in every environment. Some published screening studies explicitly note that model updates could alter outcomes. For that reason, reproducibility in AI-assisted reviews requires logging the model, version or date, interface type, prompt, and key settings. Even then, exact replay may not be possible for externally hosted models. ARISMA therefore prefers reproducibility by auditability and validation evidence over an unrealistic promise of bit-for-bit reproduction.

{\bf Reporting and ethics threat.}
Underreporting AI use is now itself a methodological threat because readers cannot assess where automation entered the workflow, whether it influenced inclusion or interpretation, or whether confidentiality, copyright, and governance issues were handled properly. This is why the joint evidence-synthesis position statement, ICMJE, COPE, and WAME all converge on disclosure and responsibility. AI cannot take authorship responsibility; the review team does.

{\bf Framework-development limitation.}
The expert consultation was purposive and formative. It was intended to identify omissions and improve the practical clarity of ARISMA rather than to establish formal consensus or demonstrate the framework's effectiveness.
The perspectives obtained may not represent all disciplines, review traditions, or institutional settings. Broader prospective evaluation of the framework therefore remains necessary.

{\bf Open methodological questions.}
Some parts of the workflow remain incompletely validated. The strongest unresolved areas are fully autonomous exclusion in high-stakes reviews, extraction of complex numeric data from tables and figures, primary-study risk-of-bias automation across designs other than standard RCT formats, and the safe use of LLMs for higher-order synthesis writing. Those are research frontiers, not mature defaults. ARISMA therefore treats them as areas for controlled experimentation, not routine deployment.

\section{Declarations}

%\textbf{Participant Consent.} 
%The expert-validation component of ARISMA involved voluntary and consented methodological feedback from academic researchers experienced in evidence synthesis and AI-assisted review workflows. 
%Participants were informed that anonymized methodological feedback and non-identifiable quotations could be incorporated into publications arising from the study.

\textbf{Expert Consultation and Consent.}
The expert-informed refinement of ARISMA involved voluntary methodological consultation with researchers and information specialists experienced in evidence synthesis. All consultees provided informed consent for participation and for the use of anonymized, non-identifiable feedback in the manuscript. No identifiable participant information is reported.

\textbf{AI Use and Authorship Responsibility.}
AI systems and large language models were discussed, analyzed, and evaluated as research subjects within this study. Any AI-assisted support used during manuscript preparation remained under full human supervision and verification. All scientific judgments, methodological decisions, interpretations, validations, and final manuscript content were reviewed, verified, and approved by the authors. No AI system is listed as an author and no AI system assumes responsibility for the work.

%\textbf{Funding Acknowledgment.} Anonymized.

%\textbf{Reproducibility and Auditability.}
%Consistent with the principles proposed in ARISMA itself, the framework emphasizes transparent reporting, validation, provenance, and human accountability for AI-assisted evidence synthesis workflows. The manuscript therefore reports the methodological assumptions, governance structure, validation rationale, and operational constraints underlying the framework in sufficient detail to support scholarly scrutiny and methodological reuse.

%****************
%{\bf Ethics.}
%This study involved feedback from academic experts on the ARISMA framework. Informed consent was obtained from all participants. Participant materials were anonymized before analysis and reporting. The manuscript reports no identifiable participant information. 

%{\bf Competing Interests.}
%The authors declare no competing interests.

%{\bf Data Availability.}
%The materials supporting the expert-validation component of this study consist of review forms and coded comments. Since these materials may contain identifiable or confidential information, they are not publicly available. 

%{\bf AI Use.}
%All scientific contributions of this work is from the authors.

%%
\section{Conclusion}

ARISMA proposes a governance-oriented framework for AI-assisted evidence synthesis grounded in validation, provenance, transparency, and human accountability. The framework does not treat AI systems as autonomous reviewers; instead, it defines the methodological conditions under which bounded AI assistance may become scientifically defensible across systematic reviews, scoping reviews, and mapping studies.
The current evidence suggests that AI can meaningfully support retrieval, prioritisation, screening assistance, extraction support, categorization, and drafting under constrained and validated conditions. At the same time, the literature consistently shows that performance remains task-dependent, unstable across contexts, and insufficient to justify unconstrained automation of consequential review decisions.
ARISMA therefore reframes AI integration as a governance problem rather than a productivity problem. The central methodological question is not whether AI can accelerate review workflows, but whether its influence on evidence selection, interpretation, and synthesis remains transparent, auditable, and scientifically controllable.
The framework’s contribution is practical as well as conceptual. By combining lifecycle guidance, validation logic, provenance structures, reporting requirements, and audit-oriented documentation, ARISMA provides review teams, peer reviewers, editors, and policymakers with a common basis for evaluating whether AI-assisted evidence synthesis remains under credible scientific oversight.
As AI systems continue to evolve, the need for transparent methodological governance will likely become more important rather than less. ARISMA is intended as a conservative foundation for that governance: one that permits responsible methodological innovation without weakening the reproducibility, interpretability, and accountability on which evidence synthesis ultimately depends. Our current work involves developing ARISMA as an open-source tool.

\bibliographystyle{splncs04}

\bibliography{bib}
%\bibliography{bib}

\end{document}